\documentclass[12pt]{article}
\usepackage[english]{babel}
\usepackage[utf8]{inputenc}
\usepackage[T1]{fontenc}

\usepackage{amsbsy}
\usepackage{amsfonts}
\usepackage{amsmath}
\usepackage{amssymb}
\usepackage{amsthm}
\usepackage{graphicx} 
\usepackage{enumerate}
\usepackage{float}
\usepackage{geometry, calc, color}
\usepackage{indentfirst}
\usepackage{longtable}
\usepackage{multirow}
\usepackage{natbib}
\usepackage{xr-hyper}
\usepackage{hyperref}
\usepackage[dvipsnames]{xcolor}
 \usepackage{booktabs}
 \usepackage{caption}
 \usepackage{subcaption}
 \usepackage[ruled]{algorithm2e}
\usepackage{mwe}
\usepackage{bm}
\usepackage{xcolor}
\usepackage{url}
\usepackage{multicol}
\usepackage{rotating}
\usepackage{fullpage}
  \usepackage{setspace}
  
\renewcommand{\arraystretch}{0.8}

\date{}

\makeatletter
\renewcommand*\env@matrix[1][\arraystretch]{%
  \edef\arraystretch{#1}%
  \hskip -\arraycolsep
  \let\@ifnextchar\new@ifnextchar
  \array{*\c@MaxMatrixCols c}}
\makeatother

\hypersetup{
  colorlinks=true,
  linkcolor=red,
  citecolor=blue,
  filecolor=red,
  urlcolor=RoyalBlue,
}
\newcommand{\blind}{1}

\protect

\begin{document}

\def\spacingset#1{\renewcommand{\baselinestretch}%
{#1}\small\normalsize} \spacingset{3}

\spacingset{1.45} 

\if1\blind
{
  \title{\bf A Bayesian latent allocation model for clustering compositional data with application to the Great Barrier Reef}
   \author{Luiza S.C. Piancastelli$^1$\footnote{Corresponding author. Email: \texttt{luiza.piancastelli@ucdconnect.ie}},\,\, Nial Friel$^{1,2}$\footnote{Email: \texttt{nial.friel@ucd.ie}} ,\,\, Julie Vercelloni$^{3}$\footnote{Email: \texttt{j.vercelloni@qut.edu.au}} ,\,\, \\ Kerrie Mengersen$^{3}$\footnote{Email: \texttt{k.mengersen@qut.edu.au}}\, and Antonietta Mira$^4$\footnote{Email: \texttt{antonietta.mira@usi.ch}}\hspace{.2cm}\\
  	{\normalsize \it $^1$School of Mathematics and Statistics, University College Dublin, Dublin, Ireland}\\
    {\normalsize \it $^2$Insight Centre for Data Analytics}\\
     {\normalsize \it $^3$School of Mathematical Sciences, Queensland University of Technology, Brisbane, Queensland, Australia}\\
     {\normalsize \it $^4$Data Science Lab, Università della  Svizzera italiana, Lugano, Switzerland}
    
    }
        
  \maketitle
} \fi
\if0\blind
{
  \bigskip
  \bigskip
  \bigskip
  \begin{center}
    {\LARGE\bf A Bayesian latent allocation model for clustering compositional data with application to the Great Barrier Reef}
\end{center}
\medskip
} \fi

\bigskip
\addtocontents{toc}{\protect\setcounter{tocdepth}{1}}

\begin{abstract}
Relative abundance is a common metric to estimate the composition of species in ecological surveys  reflecting patterns of commonness and rarity of biological assemblages. Measurements of coral reef compositions formed by four communities along Australia's Great Barrier Reef (GBR) gathered between 2012 and 2017 are the focus of this paper. We undertake the task of finding clusters of transect locations with similar community composition and investigate changes in clustering dynamics over time. During these years, an unprecedented sequence of extreme weather events (two cyclones and two coral bleaching events) impacted the 58 surveyed locations. The dependence between constituent parts of a composition presents a challenge for existing multivariate clustering approaches. In this paper, we introduce a new model that is a finite mixture of Dirichlet distributions with group-specific parameters, where cluster memberships are dictated by unobserved latent variables. The inference is carried in a Bayesian framework, where Markov Chain Monte Carlo strategies are outlined to sample from the posterior model. Simulation studies are presented to illustrate the performance of the model in a controlled setting. The application of the model to the 2012 coral reef data reveals that clusters were spatially distributed in similar ways across reefs which indicates a potential influence of wave exposure at the origin of coral reef community composition. The number of clusters estimated by the model decreased from four in 2012 to two from 2014 until 2017. Posterior probabilities of transect allocations to the same cluster substantially increase through time showing a potential homogenization of community composition across the whole GBR. The Bayesian latent allocation model highlights the diversity of coral reef community composition within a coral reef and rapid changes across large spatial scales that may contribute to undermining the future of the GBR's biodiversity.
\end{abstract}


\section{Introduction}


Species richness and relative abundance of individuals are key elements of biodiversity. Patterns of commonness and rarity are fundamental characteristics of biological assemblages used by ecologists to compare different communities and predict future developments (\cite{pacificcorals}). In this work, we deal with relative abundances of algae, hard corals, soft corals and sand in local coral communities of Australia's Great Barrier Reef (GBR) surveyed at four time points (2012, 2014, 2016 and 2017). The observed data at a reef location at a fixed time point is represented as a vector of proportions $\boldsymbol{p} = (p_1,..., p_4)$, with positive elements that sum to one. Multivariate data of this kind where information is relative is often called compositional (\cite{PawlowskyGlahn2006}). With each component comprising a part of a fixed total amount, there is inherent dependence among them and specific statistical tools are needed to handle this type of data.

The focus of this paper is on the development of statistical methodology for clustering compositional data, which is motivated by the goal of discovering groups of reef communities that share similar composition patterns. It is reasonable to assume that systems that are similar in composition were subject to common environmental conditions. Hence, by identifying transect clusters, an investigation of the mechanisms that led to the distinct composition arrangements is facilitated. Such hypothesis generating activity can help to improve the understanding of internal and external factors that impact on coral reefs locations. Moreover, it is of interest to compare composition clusters over the four years period and hence derive insights into how disturbance events affected reef community compositions. During the study period, the GBR was impacted by an unprecedented sequence of two tropical cyclones and two heat stress events. A better understanding of the effect of these multiple events is pursued as it is key to predicting future developments of these endangered ecosystems (\cite{vercelloni2020forecasting}).

For this purpose, a new model for clustering compositional data is proposed. Although the analysis of proportions or fractions is of interest in many scientific fields, statistical methodologies to deal with this type of data are less common and mostly rely on transforming the original data. In community ecology, many studies investigate how organisms utilize resources in an ecosystem, relative to their availability (\cite{valpine2013}). When referring to a habitat's occupation, such fractions are called relative abundances. Several ecological questions rely on the multivariate analysis of abundances including inference about environmental effects and species interactions (\cite{warton2015}). \cite{warton2011} emphasize how often compositional data arises from ecological problems with over one third of papers published in \textit{Ecology} over 2008 to 2009 analysing proportion data. Composition data are also of interest in epidemiology, arising from gene sequencing (\cite{gloor2016}), microbiology (\cite{gloor2017}), bioinformatics (\cite{quinn2018}) and is the core of many population studies analysis (\cite{lloyd2012}). 

A vector of $r$ compositions can be represented by $\boldsymbol{p} = (p_1,..., p_r)$, where the elements are positive ($p_i >0$ $\forall i$) and have a constant sum $k$. That is, $\sum_{i=1}^{r} p_i = k$, where $k$ is most commonly 1 (proportions) or 100 (percentages). When $k=1$ the vector $\boldsymbol{p}$ takes values in the regular unit $r$-simplex $\mathcal{S}^r$. Statistical analysis of compositional data requires acknowledging these properties, as failing to do so can lead to the problem of spurious correlations. Originally described by \cite{pearson1896}, spurious correlation arising from multivariate components that are restricted to sum to a constant has been highlighted by several authors such as \cite{chayes1960}, \cite{Vistelius1961}, \cite{Filzmoser2009}, to name a few. A recent work in this direction is by \cite{lovell2015} where a proportionality approach is proposed as an alternative to correlation for relative data. The development of appropriate statistical methodologies for compositional data analysis can be traced back to  Aitchison (\cite{Aitchison1982}, \cite{Aitchison1987}) who introduced the idea of removing the unit simplex sample space constraint through a log-ratio transformation of the data. The additive log-ratio transformation ($alt$), defined as $alt(\boldsymbol{p}) = (\log \frac{p_1}{p_r},\dots, \log \frac{p_{r-1}}{p_{r}})$, relies on a one-to-one correspondence among compositional vectors and associated log-ratio vectors (\cite{Aitchison2005}), bringing the problem to a multivariate real space. A criticism of the log-ratio analysis is the difficulty in appropriately modelling point masses at zero (\cite{fernandez2014}) and the matter of which composition to choose as divisor, something commonly done arbitrarily.
 
Alternative transformations have also been proposed in the literature. The centered log-ratio ($clr$) takes the divisor as the geometric mean of components ($\mu_G(\boldsymbol{p})$), addressing the issue of which composition to employ in the denominator. This is a popular option that has been used to analyse numerous compositional data problems. One drawback of this approach is that its resulting covariance matrix is singular (\cite{vera2011}), complicating the application of some standard statistical techniques. Another example is the isometric log-ratio ($ilr$) (\cite{Egozcue2003}). This is defined as $ilr( \boldsymbol{p}) = H \log \left( \frac{\boldsymbol{p}}{\mu_G(\boldsymbol{p})} \right)$ and transforms the $r$-simplex into an unconstrained real $r$-dimensional space (\cite{chong2018}) through the calculation of a orthonormal basis $H$. This is a mathematically elegant approach but can be complicated in practice due to the difficulty of choosing a meaningful basis (\cite{vera2011}). Some recent works in transformation based approaches for composition data are due to \cite{scealy2011} and \cite{scealy2015}. Our review of transformation methods developed for compositional data is not meant to be exhaustive, and we recommend \cite{vera2011}, \cite{vera2015}, \cite{filzmoser2018} and \cite{td2019} for more detailed overviews.

After the data are transformed, it is common to employ standard statistical analysis relying on Gaussian assumptions. Regression models employing this assumption for transformed compositions are widespread, see for example \cite{hron2012}, \cite{vander2020} and \cite{chen2016}. Drawbacks of this approach, as commented by \cite{douma2019}, are the difficulty in interpretation and possibly biased estimates. These issues frequently arise from the need to back-transform non-linear functions of the original data. Interpreting the results in the unstransformed scale is problematic due to Jensen's inequality where the further from linearity the transformation, the largest the bias. Additionally, dealing with outliers in a transformation based context is sensible. This is because it can be the case that they are accommodated by some transformation choices but not others (\cite{vidal1996}). Motivated by these difficulties, the direction we adopt there is to model the data in the original space through distributions defined on the simplex. The Beta distribution introduced in a regression context for continuous proportions by \cite{ferrari2004}, is a well-established option for modelling compositional data of two categories in the original scale. The Dirichlet distribution is a multivariate generalisation of the Beta distribution, first employed as a regression model for multivariate proportions by \cite{hijazi2009} and made readily available to practitioners by \cite{maier2014}. In ecological research, the benefit of maintaining the variables on their original scale by the application of Beta or Dirichlet models is emphasized by \cite{douma2019}. This survey encourages ecological researchers to employ Beta (2 categories) or Dirichlet (>2 categories) for data arising from continuous measurements as they provide a straightforward interpretation of parameters.

We tackle the task of clustering compositional data in the simplex, avoiding transformations and potential problems arising from it. Clustering analysis comprises a wide set of tools to identify groups of individuals in a sample such that elements belonging to the same group share similar characteristics. Different approaches to clustering exist to try to obtain meaningful partitions that highlight interesting structures in the data. For instance, two popular distance-based methods are the hierarchical agglomerative and iterative-partitioning algorithms. The first hierarchically merges similar groups in terms of pre-defined linkage criteria at each step, while the latter relocates observations until some metric is optimized. Examples of iterative-partitioning algorithms are k-nearest-neighbors (\cite{knn}) and k-means (\cite{kmeans}). The k-means algorithm has been employed to cluster transformed compositional data by \cite{gb2017}, who consider the $alt$ and $crt$ transformations.

Model-based alternatives take a statistical approach to clustering that assumes that samples arise from a mixture of probability distributions with an unknown number of components. In this context, a model is fit to explain the latent structure that generated the data. Each of the mixture components can be interpreted in some sense that corresponds to a cluster. Finite mixture models have been extensively studied in the clustering context. Early publications include \cite{edwards1965}, \cite{day1969} and \cite{wolfe}. More recent treatises include \cite{kaufman2009}, \cite{Mengersen2011}, \cite{stahl2012} and \cite{mc2016}, among others. Under model-based clustering, group assignments are performed probabilistically and it is possible to assess the uncertainty on the allocations. In addition, the number of components (or clusters) can be chosen objectively via model information criteria. According to \cite{hui2015}, model-based methods offer advantages over distance-based methods for community-level in that they can potentially provide information on the data-generating mechanism. Model-based clustering of univariate proportions was considered by \cite{Band2014} via augmented proportion density models, where inference is carried under a Bayesian framework. This model allows the inclusion of covariates but its extension to compositional vectors was not addressed. 

Our work attempts to fill this gap. We introduce a model-based approach for clustering compositional data, which comprises a finite mixture of Dirichlet distributions. Such distributions have group-specific parameters, and the cluster memberships are unobserved latent variables. Effectively, this means that we augment the original data set of $n$ observed compositions with a vector $\boldsymbol{z} = \{z_1,..., z_n\}$ of group labels, also referred to as the allocation vector.  Given $\boldsymbol{z}$, the data are conditionally independent  within latent classes. Applications of latent variable models can be found in various statistical problems such as \cite{nowocki2001} in the context of latent block models for networks, Poisson regression \cite{wedel1992}, and hidden Markov models \cite{macdonald1997}. The data augmentation approach yields a decomposition into a simpler structure that facilitates both interpretation and computations. Bayesian statistics can avail of such hierarchical specification but frequentists alternatives such as the EM algorithm are also available.

There is an emerging interest by ecologists in the Bayesian analysis of compositional data (\cite{mccarthy2007}, \cite{king2009}, \cite{korner2015}). Practical motivations in adopting this approach include the possibility of incorporating information from previous studies through the prior distribution, ready interpretation of posterior probabilities, and ease of uncertainty quantification, among others. According to \cite{Halstead2012}, hierarchical models and their analysis by Bayesian methods are increasingly common in ecology. One example is \cite{webb2018}, in which a Bayesian paradigm facilitated predictions of ecological effects in out-of-sample sites. Additionally, in \cite{webb2018}, Bayesian modelling for ecological responses to flow alterations is adopted due to the flexibility that it provides for constructing complex models. Following this growing interest and the above-mentioned benefits to practitioners, inference for the proposed clustering model is developed under the Bayesian framework, where parameters and latent variables are sampled from their full conditional distributions in a Metropolis-within-Gibbs scheme (\cite{tierney1994}). Our sampling algorithm addresses the difficulty of adopting a single proposal variance for Dirichlet parameter vectors by employing a jumping rule whereby both the median and variance depend on the current state of the Markov chain.

The introduction of our Bayesian latent allocation model fulfills our goal of underlying coral reef composition clusters under a probabilistic approach. The model fit to the 2012 ecological survey reveals the presence of four clusters and indicates the influence of wave exposure in the composition patterns, something that had not been evidenced before. A decrease in the number of clusters is observed from four in 2012 to two in 2014 onwards. Moreover, posterior group membership probabilities obtained from the Bayesian model output play an important role in revealing an increase in composition similarities or equivalently, a decreased diversity of the reef formations. Another work where clustering applied to ecology reveals insights on species distribution patterns is by \cite{clusterplants}. In this paper, the authors characterized groups of high abundance of plants in semi-deciduous forests in the Dominican Republic, identifying that patchiness and clustering in plant distributions occur at small local scales. Their findings have important implications to the spatial planning of conservation practices in the Caribbean.

The paper is organized as follows. Section \ref{data_sets} introduces the ecological survey conducted on the Great Barrier Reef that provided the compositional data that motivates the proposed methodology. Our clustering model is specified in Section \ref{model_spec}, and computation of the posterior distribution of model parameters and latent variables are outlined in Section \ref{mcmc_methods}. Section \ref{mcmc_methods} also addresses computational efficiency, the label-switching issue of mixture models and model selection. In Section \ref{simulation}, model performance in recovering groups from simulated data sets is explored. Our methodology is applied to the motivating data of coral species abundances in Section \ref{application}, where highlighting a clustering structure helped to identify spatial features of community abundance distributions and a decrease in diversity of composition throughout the years. Codes and data sets used in this paper are made available at \url{https://github.com/luizapiancastelli/Reef_Clustering}.

\section{Data sets and motivation}\label{data_sets} 
\begin{figure}
	\begin{subfigure}{.5\linewidth}
		\centering
		\includegraphics[scale=.35]{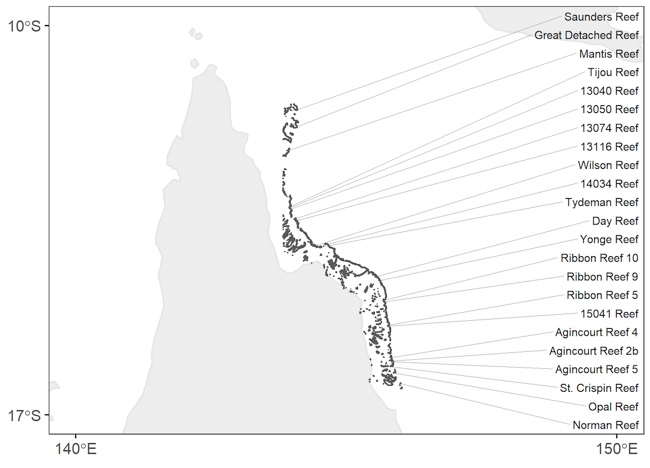}
		\caption{}
		\label{fig:sub1}
	\end{subfigure}%
	\begin{subfigure}{.5\linewidth}
		\centering
		\includegraphics[scale=.22]{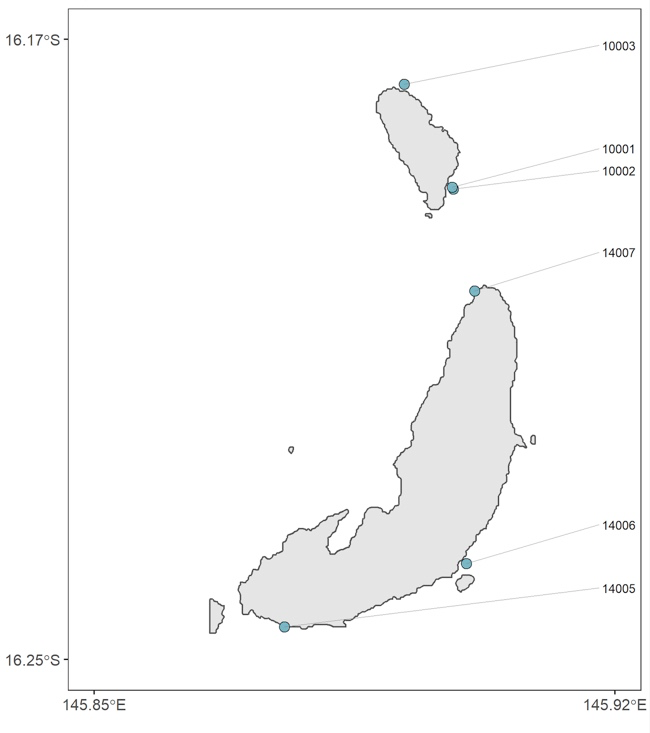}
		\caption{}
		\label{fig:sub2}
	\end{subfigure}\\[1ex]
	\begin{subfigure}{\linewidth}
		\centering
		\includegraphics[scale=.12]{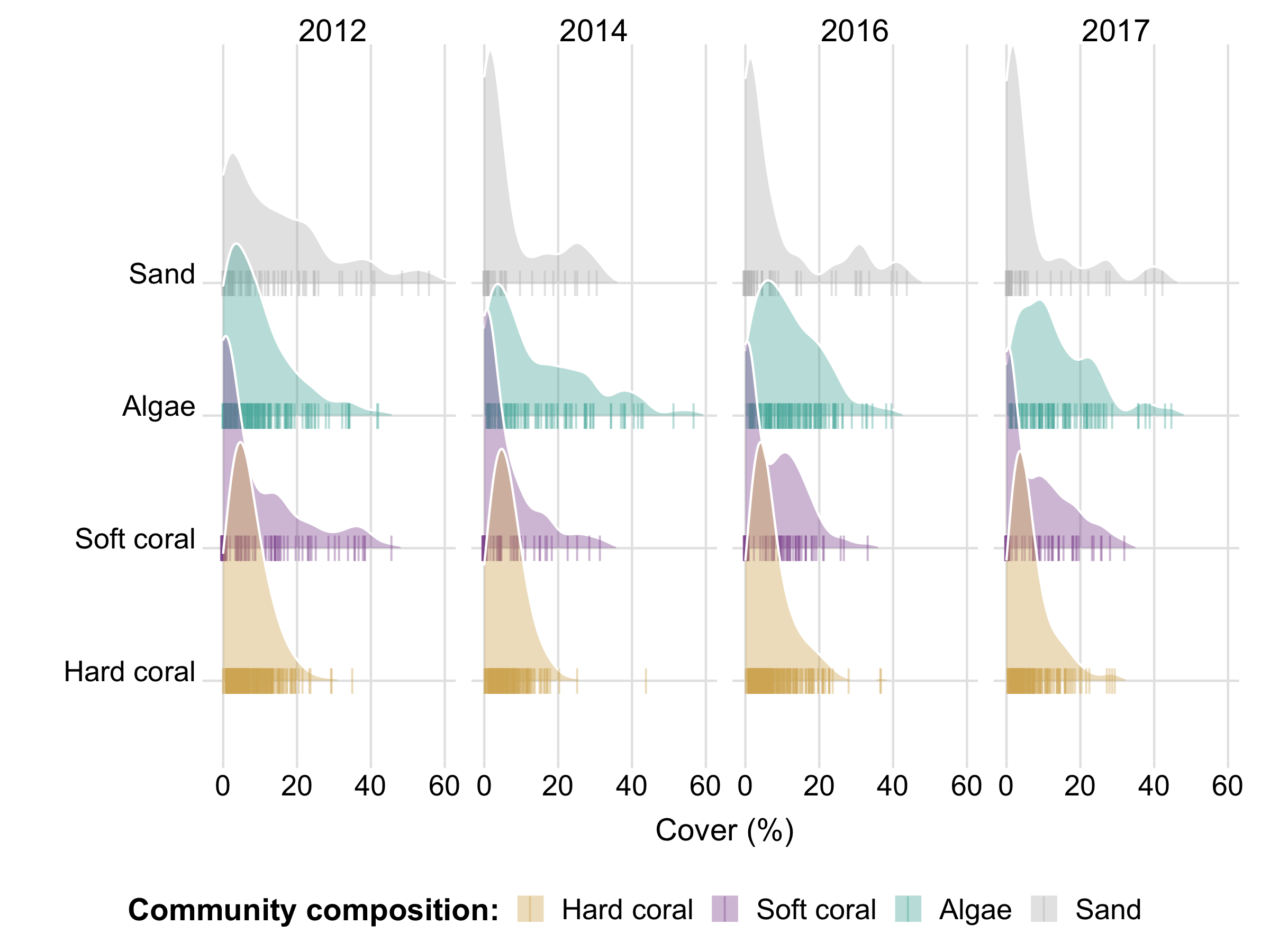}
		\caption{}
		\label{fig:sub3}
	\end{subfigure}
	\caption{(a) Locations of the surveyed coral reefs along the Great Barrier Reef. (b) Between 1-6 geo-referenced transects (blue dots) situated in different locations among a reef were variously surveyed between 2012-2017. Transects among Opal Reef are shown as example.  
	(c) Community composition of hard coral, soft coral, algae and sand were estimated using convolutional neural network from image recognition. Aggregated compositions are shown by year in (c) with bars indicating transect values.}
	\label{fig:test}
\end{figure}

The health of the Great Barrier Reef (GBR) is degrading with rapid declines of its reef-building hard coral communities already evident across broad spatial scales. During 2014-2017, the northern section of the GBR has been impacted by a series of tropical cyclones and heat stress periods that led to back-to-back mass coral bleaching \citep{vercelloni2020forecasting}. This unprecedented series of disturbances resulted in the lowest regional hard coral cover in record, shift in coral community composition and declines in coral recruitment (\cite{hughes2018global, hughes2019ecological}). While the disturbance impacts on hard corals have been well-documented, responses of other benthic communities and compositional reef structure to these declines are not well-understood.     
Ecological surveys were conducted at 23 coral reefs along the GBR between 2012-2017 (\cite{rodriguez2020contemporary}), see  Figure~\ref{fig:sub1}. Within each reef, between 1-3 transects were surveyed using a sampling methodology developed as part of the XL Seaview Catlin Survey (\cite{gonzalez2020monitoring}) as illustrated in Figure~\ref{fig:sub2}. A total of 58 transects were surveyed in 2012, and repeated in 2014 (33 transects), 2016 (49 transects) and 2017 (35 transects). Transects to resurvey were selected to capture coral reef community responses across a combination of different exposures to cyclones and coral bleaching events that impacted these reef locations between 2012 and 2017 (\cite{vercelloni2020forecasting}). Relative abundances of coral reef communities were estimated from images of the seafloor along 2 km geo-located transects of the outer reef slopes, at 10 m depth (\cite{gonzalez2014catlin, gonzalez2016scaling}). A convolutional neural network was used to automatically classify reef communities (\cite{gonzalez2020monitoring}) from a total of $\sim145,000$ images that were taken between 2012 and 2017 (\cite{rodriguez2020contemporary,vercelloni2020forecasting}). Reef communities were categorized by experts into four species, namely, hard coral, soft coral, algae, and sand and were expressed as relative coverage for each image. The compositional data of our interest displayed in Figure  \ref{fig:sub3} is the aggregation of this information across transects.

Our goal is to investigate the presence of groups of reef locations with similar community composition. If clusters are present, how many are there and what is the probability that any reef belongs to a given cluster? Community composition patterns often arise from shared environmental conditions such as common evolutionary processes or climatic events including disturbances (\cite{pacificcorals}). Therefore, the identification of clusters of coral reef community compositions can play a role in understanding the effect of known factors such as environmental disturbances or even help to uncover new ones. 

In addition, we would like to compare clusters over the years. Although a complete follow up is limited to 20 persistent observations and the years are unequally spaced, a comparison of the clusters at each time point could help to understand the effect of environmental disturbance. In 2016 and 2017, a great portion of the Great Barrier Reef was subject to back-to-back massive coral bleaching events (\cite{hughes2019ecological}). Coral bleaching is a phenomenon caused by marine heat waves, which causes corals to stress and weaken, and become subject to mortality (\cite{hughes2018global}). Coral reefs on the Great Barrier Reef were also hit by heavy winds due to categories 4 and 5 tropical cyclones in 2014 and 2015 (\cite{puotinen2016robust,vercelloni2020forecasting}). 

Adopting a Bayesian analysis to answer these questions allows to efficiently extract information and properly quantify the uncertainty behind the inference arising from small data sets. This is crucial in our context where collecting the data is time consuming and expensive. Moreover, under the Bayesian approach it is possible to complement this information with the prior knowledge of experts, if available.

\section{Model Specification}\label{model_spec}

In this section, a model for clustering compositional data in the simplex is introduced in the context of clustering coral reef locations according to their composition patterns. The proportions of four species at reef location $j$ is denoted as $\boldsymbol{p}_{j}$ for locations $j=1,\dots,n$. An unobserved variable $z_j$ represents the cluster membership of the $j^{th}$ location, assuming values in $\{1,..., k\}$, where $k$ is a pre-specified number of mixture components. The choice of $k$ is discussed in Section \ref{model_selection}. 
It is assumed that $z_j$ follows a Multinomial distribution with size one and membership probabilities $\boldsymbol{\omega} = \{\omega_1,..., \omega_k\}$. A priori, $\boldsymbol{\omega}$ is distributed according to a symmetric $k$-dimensional Dirichlet, that is, $\boldsymbol{\omega} \sim Dir(\boldsymbol{\delta})$ where $\delta_1 = \delta_2 = \cdots = \delta_k$. Hence we denote $\boldsymbol{\omega} \sim Dir({\delta})$. Conditionally on the collection of unobserved allocations $\boldsymbol{z} \equiv \{z_1, \cdots, z_n \}$, the likelihood model assumes a simple form, dictating that each $\boldsymbol{p_{j}}$ arises from a $k$-component mixture of Dirichlet distributions with cluster-specific parameters. This is denoted as $\boldsymbol{p}_{j}|z_j  = l  \sim Dir(\boldsymbol{\rho}_l)$, where $\boldsymbol{\rho}_l = (\rho_{l1},, \dots, \rho_{l4})$. The complete set of parameters is represented as $\underline{\boldsymbol{\rho}}$, a ($k \times 4$) matrix with elements $\rho_{li} > 0$, for $l=1,..,k$ and $i=1,...,4$. Under the assumption of local independence, the conditional data likelihood for a set of $n$ reef locations is written as

\begin{eqnarray*}\label{likelihood}
f(\underline{\boldsymbol{p}}| \boldsymbol{z}, \underline{\boldsymbol{\rho}}) = \prod_{j=1}^{n} \left( \frac{\Gamma(\sum_{i=1}^{4}  \rho_{z_j,i})}{\prod_{i=1}^{4} \Gamma(\rho_{z_j, i})} \prod_{i=1}^{4} p_{ij}^{\rho_{z_j, i} - 1} \right),
\end{eqnarray*}
where $\underline{\boldsymbol{p}}$ is a matrix with entries $\boldsymbol{p}_{j}$. Cluster-specific Dirichlet parameters can be used to interpret the group species abundance patterns considering the following:
 
\begin{enumerate}
	\item The parameter vector normalised by its sum, i.e., $\rho_{li}/\sum_{i=1}^{4}\rho_{li}$ gives us the mean species abundances in cluster $l$.
	\item $\sum_{i=1}^{4}\rho_{li}$ relates to the variability of the distribution, where lower values are associated to a higher variability.
\end{enumerate}

A hierarchical prior is set for the $\underline{\boldsymbol{\rho}}$ elements which are independent and Gamma($\alpha$, $\beta$) distributed. The introduction of the hyperpriors with fixed hyperparameters $\alpha \sim Exp(\gamma)$ and $\beta \sim \mbox{Gamma}(\phi, \lambda)$ model our uncertainty on the $\alpha$ and $\beta$ values. Under these assumptions, the posterior model is specified by the product

\begin{eqnarray*}
	\pi(\alpha, \beta, \boldsymbol{\omega},  \underline{\boldsymbol{\rho}}, \boldsymbol{z}| \underline{\boldsymbol{p}}) \propto f(\underline{\boldsymbol{p}}| \boldsymbol{z},  \underline{\boldsymbol{\rho}}) \pi( \underline{\boldsymbol{\rho}}| \alpha, \beta) \pi( \boldsymbol{z}|\boldsymbol{\omega}) \pi(\alpha|\gamma) \pi(\beta| \phi, \lambda) \pi(\boldsymbol{\omega}|\delta).
\end{eqnarray*}

Leveraging the Multinomial-Dirichlet conjugacy, we can integrate out $\boldsymbol{\omega}$, yielding the partially collapsed posterior model

\begin{eqnarray}\label{posterior}
\pi(\alpha, \beta,  \underline{\boldsymbol{\rho}}, \boldsymbol{z}| \underline{\boldsymbol{p}}) \propto f(\underline{\boldsymbol{p}}| \boldsymbol{z},  \underline{\boldsymbol{\rho}}) \pi( \underline{\boldsymbol{\rho}}| \alpha, \beta) \pi( \boldsymbol{z}|\delta) \pi(\alpha|\gamma) \pi(\beta| \phi, \lambda),
\end{eqnarray}
where $\pi(\boldsymbol{z}|\delta) \propto \frac{\prod_{l=1}^{k} \Gamma(a_l)}{\Gamma(\sum_{l=1}^{k} a_l)}$, with  $a_l = \sum_{j=1}^{n} I(z_j =l) + \delta$.

We propose to make inference by sampling from (\ref{posterior}) and we develop algorithms to draw from the posterior distribution of model parameters and latent variables in the next section.

\section{MCMC methods}\label{mcmc_methods}

A Metropolis-within-Gibbs scheme is presented in this section, where model parameters and latent variables are drawn from their conditional distributions. 

\subsection{Gibbs sampling update of latent allocations}

First, the entries of the cluster membership vector $\boldsymbol{z}$ are updated one element at a time drawing from the conditional distribution of $z_j$ given $\boldsymbol{z}_{-j}$,  where $\boldsymbol{z}_{-j}$ denotes the allocation vector without the $j^{th}$ observation. Our goal is to simulate from $\pi(z_j| \boldsymbol{z}_{-j},  \boldsymbol{\rho}_{z_j}, \delta, \boldsymbol{p}_{j})$, which is proportional to

\begin{eqnarray}
\pi(z_j| \boldsymbol{z}_{-j}, \boldsymbol{\rho}_{z_j}, \delta,  \boldsymbol{p}_{j}) \propto f(\boldsymbol{p}_{j}|z_j, \boldsymbol{\rho}_{z_j}) \pi(\boldsymbol{z}|\delta).\label{conditional_zj}
\end{eqnarray}

As the target distribution of interest is not of known form, a Metropolis update can be employed. For each $j$, a new allocation $z_j'$ is drawn from a discrete Uniform$\{1, \dots, k\}$ kernel and accepted or rejected with the probability that preserves detailed balance with respect to the full conditional distribution in (\ref{conditional_zj}). Given the current states of $\boldsymbol{z}$ and $\boldsymbol{\underline{\rho}}$, $z_j'$ is accepted with probability $\min \left\{ 1, \frac{\pi(z_j'|\boldsymbol{z}_{-j},  \boldsymbol{\rho}_{z_j}, \delta, \boldsymbol{p}_{j})}{\pi(z_j|\boldsymbol{z}_{-j},  \boldsymbol{\rho}_{z_j}, \delta, \boldsymbol{p}_{j})} \right\}$ where $\pi(\cdot|\boldsymbol{z}_{-j},  \boldsymbol{\rho}_{z_j}, \delta, \boldsymbol{p}_{j})$ is given in \ref{conditional_zj}. The Metropolis update is concluded after all elements of $\boldsymbol{z}$ have been visited. Another possibility is to consider a Gibbs sweep. This alternative requires that we calculate for every observation its conditional posterior probability of belonging to cluster $l$, for $l=1,...,k$. This is denoted by $m_{il}$ and given by

\begin{eqnarray}\label{gibbs}
m_{il} = \pi(z_j = l | \boldsymbol{z}_{-j}, \boldsymbol{\rho}_{z_j}, \delta,  \boldsymbol{p}_{j}) = \frac{\pi(z_j = l| \boldsymbol{z}_{-j},  \boldsymbol{\rho}_{z_j}, \delta, \boldsymbol{p}_{j})} { \sum_{l=1}^{k} \pi(z_j = l| \boldsymbol{z}_{-j},  \boldsymbol{\rho}_{z_j}, \delta, \boldsymbol{p}_{j}) }.
\end{eqnarray}

Our strategy is to start with $j=1$ and update the observations sequentially but the processing order does not matter in the limit. The Gibbs sweep requires that, for every $j$, conditional posterior probabilities in (\ref{gibbs}) are calculated for $l=1,...,k$. Hence, the number of function evaluations grows with the number of observations and clusters, making the Metropolis update to be preferred in cases where $n$ and $k$ are large. In our application to the coral species abundances data $n$ and $k$ are small, so the Gibbs step is taken.

\subsection{Metropolis-Hastings update of $\alpha$}
The second step in our MCMC scheme is to update $\alpha$. The expression of the conditional distribution of this parameter up to proportionality is

\begin{eqnarray}\label{target_alpha}
\pi(\alpha|\beta, \underline{\boldsymbol{\rho}}, \gamma) \propto \frac{\beta^{4k\alpha}}{\Gamma(\alpha)^{4k}} \left(  \prod_{l=1}^{k} \prod_{i=1}^{4} \rho_{li}^{\alpha -1 } \right) e^{-\gamma \alpha}.
\end{eqnarray}

We propose to draw from (\ref{target_alpha}) using a Metropolis-Hastings algorithm with a log-Normal jumping distribution. Denoting by $\alpha^{(t)}$ the current state of the chain, a new state $\alpha^*$ is simulated according to a log-Normal($\log(\alpha^{(t)}), \sigma_\alpha$) and accepted or rejected with the proper probability. This proposal distribution is not symmetric, having median equal to the current state $\alpha^{(t)}$ and $\sigma_\alpha$ calibrating the proposal variance. Accordingly, the ratio $\frac{q(\alpha^*, \alpha^{(t)})}{q(\alpha^{(t)}, \alpha^{*})}$ need to be accounted in the Metropolis-Hastings acceptance, where $q(\alpha',\alpha^\dagger)$ denotes the probability of moving to state $\alpha^\dagger$ from $\alpha'$. Given that $q(\alpha', \alpha^\dagger) \equiv \mbox{log-Normal}(\alpha^\dagger|\log(\alpha'), \sigma_\alpha)$, the proposal ratio simplifies to $\frac{\alpha^*}{\alpha^{(t)}}$. Pseudocode in Algorithm \ref{mh_alpha} describes the steps to update $\alpha$.

\vspace{0.5cm}
\begin{algorithm}[H]
\small
	\caption{Metropolis-Hastings step for $\alpha$}\label{mh_alpha}
	
	1. Propose a new value  $\alpha^*$ drawing from a  log-Normal($\alpha^{(t)}$, $\sigma_{\alpha}$). \\
	
	2. Draw $u \sim U(0,1)$.
	
	3. Calculate $r(\alpha^{(t)}, \alpha^*) = \mbox{min} \left( 1,\frac{\pi(\alpha^*|...) \alpha^*}{\pi(\alpha^{(t)}|...)\alpha^{(t)}} \right)$.
	
	\eIf{u < $r(\alpha^{(t)}, \alpha^*)$}{
		$\alpha^{(t+1)} = \alpha^*$
	}{
		$\alpha^{(t+1)} = \alpha^{(t)}$
	}
\end{algorithm}

\subsection{Gibbs update of $\beta$}

Next, given the current states of $\alpha$ and $\boldsymbol{\underline{\rho}}$ plus the hyperprior parameters ($\phi$, $\lambda$), we update $\beta$ from its conditional distribution

\begin{eqnarray}
\pi(\beta| \alpha,  \underline{\boldsymbol{\rho}}, \phi, \lambda) \propto \beta^{4\alpha k + \phi -1} \exp \left\{  -\beta   \left( \lambda + \sum_{l=1}^{k}  \sum_{i=1}^{4}  \rho_{li} \right)   \right\},
\end{eqnarray}
which we recognize as a Gamma distribution with shape $\phi + 4\alpha k$ and rate $\lambda + \sum_{l=1}^{k}  \sum_{i=1}^{4}  \rho_{li}$. Hence, a Gibbs update is taken for this parameter simulating $\beta^{(t+1)} \sim \mbox{Gamma}(\phi + 4\alpha^{(t+1)} k, \lambda + \sum_{l=1}^{k}  \sum_{i=1}^{4}  \rho_{li}^{(t)})$.

\subsection{Metropolis-Hastings update of Dirichlet parameters}

Our MCMC scheme is concluded with the update of the Dirichlet parameters given the current values of $\boldsymbol{z}$, $\alpha$ and $\beta$. For a fixed cluster $l$, each element of the group-specific vector $\boldsymbol{\rho}_l$ will be updated given the others by drawing from the conditional distribution $\pi(\rho_{li}| \boldsymbol{\rho}_{l-i}, \alpha, \beta, \underline{\boldsymbol{p}})$, with $ \boldsymbol{\rho}_{l-i}$ denoting $\boldsymbol{\rho_l}$ without the $i^{th}$ element. We can write

\begin{eqnarray}
\pi(\rho_{li}| \boldsymbol{\rho}_{l-i}, \alpha, \beta, \underline{\boldsymbol{p}}) \propto  \prod_{j: z_j = l}  \left(\frac{\Gamma(\sum_{i=1}^{4} \rho_{li})}{\prod_{i=i}^{4} \Gamma(\rho_{li})}   \prod_{i=1}^{4} \rho_{li}^{\alpha -1}  \right) \prod_{i=1}^{4} \rho_{li}^{\alpha-1} e^{-\beta \sum_{i=1}^{4} \rho_{li}}.
\end{eqnarray}

Since this is not of known form, we set up a Metropolis-Hasting update for $\boldsymbol{\underline{\rho}}$. One distinction from Algorithm \ref{mh_alpha} is that it is now unfeasible to adopt a single proposal variance parameter. This is due to the fact that the $\boldsymbol{\underline{\rho}}$ matrix can contain elements of various magnitudes, where it is unlikely that we can control the Markov chain acceptance rate for all $\rho_{li}, i=1,\cdots,4; l=1,\cdots,k$ through a single parameter. Naturally, one option is to consider element-wise proposals $\rho_{li}' \sim \mbox{log-Normal}(\log(\rho_{li}), \sigma_{li})$ but this results in an increasing number of hyper-parameters to tune.

We address this by introducing a proposal variance that depends on the current state. Denoting by $p_{var}$ a percentage of the current $\rho_{li}^{(t)}$, a proposed state $\rho_{li}^{*}$ is drawn with median $\rho_{li}^{(t)}$ and variance $p_{var} \rho_{li}^{(t)}$. To achieve this, we set the log-Normal$(\log(\rho_{li}),\sigma)$ variance to be equal to $p_{var} \rho_{li}^{(t)}$. We then obtain the $\sigma$ that yields the target variance by solving $(\exp(\sigma^2)-1)\exp(2\log(\rho_{li}^{(t)})+\sigma^2) = p_{var} \rho_{li}$. We denote by $\sigma(\rho_{li}^{(t)})$ the solution to this equation which is analytical and equal to 

\begin{eqnarray}\label{eq:sigma}
\sigma(\rho_{li}^{(t)}) = \log \sqrt{\frac{ \sqrt{4 (p_{var} \rho_{li}^{(t)}) \rho_{li}^{{(t)}^2} + \rho_{li}^{{(t)}^4}} + \rho_{li}^{{(t)}^2}}{2\rho_{li}^{{(t)}^2}}}.
\end{eqnarray}

The jumping distribution $q(\rho_{li}', \rho_{li}^\dagger)$, a log-Normal$(\rho_{li}^\dagger|\log(\rho_{li}'), \sigma(\rho_{li}') )$, is accounted for in the acceptance ratio of the algorithm that includes the ratio of transition probabilities. In line with many Metropolis-Hastings algorithms, we recommend that $p_{var}$ is calibrated in pilot runs by monitoring the acceptance rate. An acceptance rate of approaching 44\% is commonly pursued for univariate the random walk Metropolis-Hastings (\cite{gelman1995}). These concepts are introduced in Algorithm \ref{MH_rho_adapt}, where the steps to update the entire $\underline{\boldsymbol{\rho}}$ parameter matrix are described with pseudocode.

	\begin{algorithm}[]
	\small
	Start with $l=1$.
	
	\While{l <= k}{
		
		Set $i=1$. \\
		\While{i <= 4}{
			
			Calculate $\sigma(\rho_{li}^{(t)})$ using (\ref{eq:sigma}).
			
			Propose $\rho_{li}^* \sim \mbox{log-Normal}(\log({\rho_{li}^{(t)}}), \sigma(\rho_{li}^{(t)}))$.
			
			Calculate $\sigma(\rho_{li}^{*})$ using (\ref{eq:sigma}).
			
			Draw $u \sim U(0,1)$.
			
			Calculate $r(\rho_{li}^{(t)}, \rho_{li}^*) = \mbox{min} \left( 1,\frac{\pi(\boldsymbol{\rho}_{l}^*|...) q(\rho_{li}^*, \rho_{li}^{(t)}) }{\pi(\boldsymbol{\rho}_{l}^{(t)}|...)q(\rho_{li}^{(t)}, \rho_{li}^*)} \right)$, \\
			where $q(\rho_{li}^{(t)}, \rho_{il}^{*}) \sim$ log-Normal$(\rho_{li}^{(t)}, \sigma(\rho_{li}^{(t)}) )$.

			\eIf{u < $r(\rho_{li}^{(t)}, \rho_{li}^*)$}{
				
				$\rho_{li}^{(t)} = \rho_{li}^*$
				
			}{
				$\rho_{li}^{(t)} = \rho_{li}^{(t)}$
			}
			$i = i+1 $
		}
		$k = k +1$
		
	}
	\caption{Metropolis-Hastings algorithm for $\underline{\boldsymbol{\rho}}$}\label{MH_rho_adapt}
\end{algorithm}	

This concludes an MCMC scheme to draw from all the model parameters and latent variables. Inference is carried by running the chain with above steps until convergence to its stationary distribution is achieved. The simulated draws are treated as a random sample from the posterior model and used to obtain summaries of interest on cluster allocations and group-specific Dirichlet parameters. Before this is done, post-processing the resulting chain is in order. This is because the model posterior is invariant to the permutation of group labels, resulting in the well-known label-switching issue. We discuss how to circumvent this and other computational aspects in the next sections.

\subsection{Computational efficiency}

Increasing the number of parameters (either the number of clusters or $n$), increases the computational cost of running the MCMC algorithm. Computational efficiency was tackled by implementing the proposed MCMC scheme in Python, increasing its speed drastically through the usage of the Numba (\cite{numba}) library.  Numba is a just-in-time compiler that translates Python code into machine code, compiling the functions without the involvement of the Python interpreter. This approach reduced the original computation time by 99.9\%.

\subsection{Label switching}
The model proposed in this work is prone to the label-switching issue, a well-known and frequently encountered phenomenon in the Bayesian estimation of finite mixture or hidden Markov models (\cite{redner1984}, \cite{diebolt1994}, \cite{jasra2005}). This term is used to refer to the invariance of the likelihood to the relabelling of components in the mixture. We tackle this by post-processing our MCMC chains with a relabelling algorithm. We employ Stephen's methodology (\cite{stephens}) which makes use of the classification probabilities in (\ref{gibbs}). Let $M(\boldsymbol{\underline{\rho}})$ denote the $n$ x $k$ matrix of classification probabilities that is saved for every MCMC iteration. In a decision theoretic approach, the labels are permuted so to agree as much as possible with $M(\boldsymbol{\underline{\rho}})$, according to the Kullback-Leibler divergence.  If a Metropolis-Hasting update is taken and $M(\boldsymbol{\underline{\rho}})$ is not available, other strategies can be used, such as the data-based relabelling procedure of \cite{rodriguez2014}. After relabelling, the posterior sample can be used to make inference on the cluster allocations and group-specific parameters.

\subsection{Cluster entropy}

We now introduce the idea of using entropy as a measure of within-cluster variability. The entropy of a random variable is a measure describing the average level of information related to its possible outcomes, where maximum entropy corresponds to maximum uncertainty. A
continuous random variable $X$ with probability density $f(\cdot)$ and support $\chi$ has continuous (or differential) entropy defined by $H(X) = - \int_\chi f(x) \log f(x) d x$. If $f(\cdot)$ is a Dirichlet distribution with parameter vector $\boldsymbol{\rho} = (\rho_1, ..., \rho_4)$, there is an analytical expression for $H(X)$ given by $H(X) = \log \beta(\boldsymbol{\rho}) + (\rho_0 -4)\psi(\rho_0) - \sum_{i=1}^4 (\rho_i -1)\psi(\rho_i)$, where $\rho_0=\sum_{i=1}^4 \rho_i$, $\beta(\boldsymbol{\rho})=\frac{\prod_{i=1}^4 \Gamma(\rho_i)}{\Gamma(\rho_0)}$,  and $\psi$ denotes the digamma function. Draws from the cluster-specific parameters from our Bayesian model output can be used to calculate the posterior entropy of the clusters. This provides an easily interpretable numerical quantification of intra-cluster diversity that can be used to contrast the identified groups.

\section{Simulation studies}\label{simulation}

\subsection{Model estimation}

In this section, simulation studies are conducted to assess the ability of our model to recover clusters from data simulated according to the likelihood model. To mimic our coral reefs application, mixtures of four-dimensional Dirichlet distributions are considered as well as small total sample sizes of 30 or 50 observations. The observations are evenly (or approximately) distributed among either two or three clusters. In each case, different settings of $\boldsymbol{\underline{\rho}}$ are chosen to correspond to different levels of group separation, resulting in six simulated data sets. Group separation is quantified according to the pairwise Hellinger distance between the mixture components, and the scenarios are denoted as high, moderate, or low separation.

The Hellinger distance provides a similarity quantification among two probability distributions, assuming values in [0,1], where smaller values indicate more similar distributions. The squared Hellinger's distance between probability functions $f$ and $g$ is defined by $H^2(f,g) = 1 - \int \sqrt{\frac{f(x)}{g(x)}} g(x) d x$, where the integral can be viewed as the expectation $E_g \sqrt{\frac{f(x)}{g(x)}}$. With $m$ independent draws from $g()$, we can approximate the squared Hellinger's distance with a Monte Carlo average $\widehat{H}^2(f,g) = 1 - \frac{1}{m} \sum_{i=1}^{m} \sqrt{\frac{f(x_i)}{g(x_i)}}$. Since the number of observations is not increased with the number of clusters, it is expected that the support for distribution overlap diminishes as fewer observations per group become available. The joint maximum a posteriori (MAP) estimators ($\widetilde{\boldsymbol{z}}, \widetilde{\boldsymbol{\underline{\rho}}}$) are reported and  $\widetilde{\boldsymbol{z}}$ is compared to the true group allocations used to generate the data.

\begin{table}[]
\centering
\scriptsize
\begin{tabular}{@{}cccccc@{}}
\toprule
\textbf{Separation}       & \textbf{\begin{tabular}[c]{@{}c@{}}Hellinger's\\ distance\end{tabular}}                                   & \textbf{\begin{tabular}[c]{@{}c@{}}True\\ parameters\end{tabular}}                                                                            & $\mathbf{n}$ & \textbf{\begin{tabular}[c]{@{}c@{}}MAP\\ parameters\end{tabular}}                                                                                                                     & \textbf{\begin{tabular}[c]{@{}c@{}}Confusion \\ matrix\end{tabular}}                                                     \\ \midrule
\multirow{2}{*}{High}     & \multirow{2}{*}{1}                                                                                        & \multirow{2}{*}{\begin{tabular}[l]{@{}c@{}}$\boldsymbol{\rho}_1= (15, 15, 1, 1)$\\ $\boldsymbol{\rho}_2 = (2, 2, 15, 20)$\end{tabular}}                                 & 30         & \begin{tabular}[l]{@{}c@{}}$\boldsymbol{\widetilde{\rho}}_1 = (15.2, 16.0, 1.1, 1.3)$\\ $\boldsymbol{\widetilde{\rho}}_2 = (1.6, 2.2, 12.5, 18.6)$\end{tabular}                                                 & \begin{tabular}[c]{@{}c@{}}$\begin{bmatrix}[0.3]\\ 15 & 0 \\\\ 0 & 15\\  \end{bmatrix}$ \vspace{0.1cm} \end{tabular}                             \\
                          &                                                                                                           &                                                                                                                                               & 50         & \begin{tabular}[l]{@{}c@{}}$\boldsymbol{\widetilde{\rho}}_1 = (17.5, 17.1, 1.3, 1.6)$\\ $\boldsymbol{\widetilde{\rho}}_2 = (2.2, 3.0, 20.8, 22.2)$\end{tabular}                                                 & \begin{tabular}[c]{@{}c@{}}$\begin{bmatrix}[0.3]\\ 25 & 0 \\\\ 0 & 25\\ \end{bmatrix}$\end{tabular}                       \vspace{0.5cm}      \\ 
\multirow{2}{*}{Moderate} & \multirow{2}{*}{0.75}                                                                                     & \multirow{2}{*}{\begin{tabular}[l]{@{}l@{}}$\boldsymbol{\rho}_1= (10, 9, 3, 2)$\\ $\boldsymbol{\rho}_2 = (10, 8, 5, 7)$\end{tabular}}                                   & 30         & \begin{tabular}[l]{@{}l@{}}$\boldsymbol{\widetilde{\rho}}_1 = (11.9, 10.3, 4.2, 3.3)$ \\  $\boldsymbol{\widetilde{\rho}}_2 = (9.8,  9.2, 7.3, 11.0)$\end{tabular}                                                 & \begin{tabular}[c]{@{}c@{}}$\begin{bmatrix}[0.3]\\  15 & 0 \\\\  6  & 9 \\ \end{bmatrix} \vspace{0.1cm} $\end{tabular}                          \\
                          &                                                                                                           &                                                                                                                                               & 50         & \begin{tabular}[l]{@{}l@{}}$\boldsymbol{\widetilde{\rho}}_1 = (15.6, 11.2,  4.1,  3.5)$\\ $\boldsymbol{\widetilde{\rho}}_2 = (11.5, 8.4, 6.8, 8.4)$\end{tabular}                                                & \begin{tabular}[c]{@{}c@{}}$\begin{bmatrix}[0.3]\\  22 & 0 \\\\  3 & 22 \\ \end{bmatrix}$\end{tabular}                 \vspace{0.5cm}         \\
\multirow{2}{*}{Low}      & \multirow{2}{*}{0.45}                                                                                     & \multirow{2}{*}{\begin{tabular}[l]{@{}l@{}}$\boldsymbol{\rho}_1= (9, 8, 4, 5)$\\ $\boldsymbol{\rho}_2 = (13, 12, 6, 12)$\end{tabular}}                                  & 30         & \begin{tabular}[l]{@{}l@{}}$\boldsymbol{\widetilde{\rho}}_1 = (15.0, 9.9, 14.3, 8.9)$\\ $\boldsymbol{\widetilde{\rho}}_2 = (13.1, 12.5, 6.6, 9.0)$\end{tabular}                                                 & \begin{tabular}[c]{@{}c@{}}$\begin{bmatrix}[0.3]\\ 1  & 14 \\\\ 0 & 15\\ \end{bmatrix}$ \vspace{0.1cm}\end{tabular}                            \\
                          &                                                                                                           &                                                                                                                                               & 50         & \begin{tabular}[l]{@{}l@{}} $\boldsymbol{\widetilde{\rho}}_1 = (18.8, 12.3, 11.5, 8.2)$ \\ $\boldsymbol{\widetilde{\rho}}_2 = (13.8, 13.2, 4.9, 10.2)$ \end{tabular}                                               & \begin{tabular}[c]{@{}c@{}}$\begin{bmatrix}[0.3]\\ 8  & 17  \\\\ 1 & 24 \\ \end{bmatrix}$\end{tabular}                    \\
                          
                          \midrule

\multirow{2}{*}{High}     & \multirow{2}{*}{\begin{tabular}[c]{@{}c@{}} $\widehat{H}$(1, 2): 1\\ $\widehat{H}$(1, 3): 1\\ $\widehat{H}$(2, 3): 1\end{tabular}}          & \multirow{2}{*}{\begin{tabular}[l]{@{}l@{}}$\boldsymbol{\rho}_1= (10, 10, 10, 10)$\\ $\boldsymbol{\rho}_2 = (1, 2, 15, 18)$\\ $\boldsymbol{\rho}_3 = (10, 12, 1, 0.5)$\end{tabular}} & 30         & \begin{tabular}[l]{@{}l@{}} $\boldsymbol{\widetilde{\rho}}_1 = (11.7, 14.2, 10.7, 10)$ \\ $\boldsymbol{\widetilde{\rho}}_2 = (2.2, 2.5, 19.3, 2.2)$\\ $\boldsymbol{\widetilde{\rho}}_3 = (9.8, 12.7, 1.4, 1.9)$\end{tabular} & \begin{tabular}[c]{@{}c@{}}$\begin{bmatrix}[0.5]\\ 10 & 0 & 0 \\\\ 0 & 10  & 0\\ 0 & 0 & 10\\ \end{bmatrix}$\end{tabular}  \vspace{0.1cm}   \\ 
                          &                                                                                                           &                                                                                                                                               & 50         & \begin{tabular}[c]{@{}c@{}}$\boldsymbol{\widetilde{\rho}}_1 = (11.2, 12.5, 11.6, 15)$\\ $\boldsymbol{\widetilde{\rho}}_2 = (1.6,  3.5, 20.4, 21.1)$  \\  $\boldsymbol{\widetilde{\rho}}_3 = (14.4, 15.3,  1.4,  0.7)$\end{tabular}     & \begin{tabular}[c]{@{}c@{}}$\begin{bmatrix}[0.4]\\ 16 & 0 & 0 \\\\   0 & 17  & 0 \\\\ 0 & 0 & 17\\ \end{bmatrix}$ \end{tabular} \vspace{0.5cm}\\
\multirow{2}{*}{Moderate} & \multirow{2}{*}{\begin{tabular}[c]{@{}c@{}}$\widehat{H}$(1, 2): 1\\ $\widehat{H}$(1, 3): 1\\ $\widehat{H}$(2, 3): 0.72\end{tabular}}       & \multirow{2}{*}{\begin{tabular}[l]{@{}l@{}}$\boldsymbol{\rho}_1= (1, 5, 1, 15)$\\ $\boldsymbol{\rho}_2 = (13, 7, 0.5, 4)$\\ $\boldsymbol{\rho}_3 = (9, 8, 1, 1)$\end{tabular}}       & 30         & \begin{tabular}[l]{@{}l@{}}$\boldsymbol{\widetilde{\rho}}_1 = (1.4, 6.1, 1.5, 13.7)$\\  $\boldsymbol{\widetilde{\rho}}_2 = (9.7, 6.4, 0.7, 1.7)$ \\ $\boldsymbol{\widetilde{\rho}}_3 = (5.5, 2.8, 11.4, 3.2)$\end{tabular}     & \begin{tabular}[c]{@{}c@{}}$\begin{bmatrix}[0.4]\\  10 & 0 & 0  \\\\ 0 & 10 & 0 \\\\ 0 & 10 & 0\\ \end{bmatrix}$\end{tabular}               \\
                          &                                                                                                           &                                                                                                                                               & 50         & \begin{tabular}[c]{@{}c@{}}$\widetilde{\boldsymbol{\rho}}_1 = (1.6, 3.4, 1.0, 12.3)$\\ $\boldsymbol{\widetilde{\rho}}_2 = (11.9, 8.0, 0.8, 4.1)$ \\ $\boldsymbol{\widetilde{\rho}}_3 = (12.9, 9.9, 1.4, 0.8)$ \end{tabular}    & \begin{tabular}[c]{@{}c@{}}$\begin{bmatrix}[0.5]\\  15 & 0 & 1\\\\ 0 & 17  & 0 \\\\ 0 & 6 & 11\\ \end{bmatrix}$ \vspace{0.5cm}\end{tabular}     \\
\multirow{2}{*}{Low}      & \multirow{2}{*}{\begin{tabular}[c]{@{}c@{}}$\widehat{H}$(1, 2): 0.68\\ $\widehat{H}$(1, 3): 0.56\\ $\widehat{H}$(2, 3): 0.69\end{tabular}} & \multirow{2}{*}{\begin{tabular}[c]{@{}c@{}}$\boldsymbol{\rho}_1= (6, 9, 1, 1)$\\ $\boldsymbol{\rho}_2 = (8, 8, 3, 3)$\\ $\boldsymbol{\rho}_3 = (7, 7, 4, 1)$\end{tabular}}           & 30         & \begin{tabular}[l]{@{}l@{}} $\boldsymbol{\widetilde{\rho}}_1 = (5.9, 5.7, 1.9, 1.2)$ \\ $\boldsymbol{\widetilde{\rho}}_2 = (10.4, 8.4, 9.0, 5.9) $ \\ $\widetilde{\boldsymbol{\rho}}_3 = (5.4, 3.2, 7.7, 5.9) $  \end{tabular}      & \begin{tabular}[c]{@{}c@{}}$\begin{bmatrix}[0.4]\\  10 & 0 & 0 \\\\ 9 & 1 & 0 \\\\ 10 & 0 & 0\\ \end{bmatrix}$\end{tabular}              \vspace{0.1cm}  \\
                          &                                                                                                           &                                                                                                                                               & 50         & \begin{tabular}[l]{@{}l@{}} $\boldsymbol{\widetilde{\rho}}_1 = (7.4, 13.9,  1.5,  1.1)$\\$\widetilde{\boldsymbol{\rho}}_2 = (10.5, 10.7,  2.9,  4.1)$ \\ $\widetilde{\boldsymbol{\rho}}_3 = (17.6,  9.7, 12.1,  2.6)$\end{tabular}      & \begin{tabular}[c]{@{}c@{}}$\begin{bmatrix}[0.4]\\  15 & 1 & 0 \\\\ 2 & 13  & 2 \\\\ 1 & 7 & 9\\ \end{bmatrix}$\end{tabular}    \\ 
                        \bottomrule
                        \\
\end{tabular}
	\caption{Results of a simulation study assessing our model's ability to recover cluster partitions from simulated data. Three settings named high, moderate and low separation levels are considered for two (top-half of the table) and three (bottom-half of the table) component Dirichlet mixtures. Distribution overlap is quantified via the Hellinger distance (HD) among pairs where this is increased over the 2-cluster scenarios. The study design under 3 clusters differs in that the moderate case introduces similarity between two of the groups while there is distribution overlap among the three pairs in the low setting. Under each parameter configuration, the sample sizes of 30 and 50 observations are investigated. The MAP pair $(\widetilde{\boldsymbol{z}} , \widetilde{\underline{\boldsymbol{\rho}}})$ is reported as well as the confusion matrix between the true allocations used to generate the data (rows) and the MAP partition (columns).}\label{table_sim}
\end{table}

\begin{figure} 
	\includegraphics[width=.32\textwidth,height=7cm]{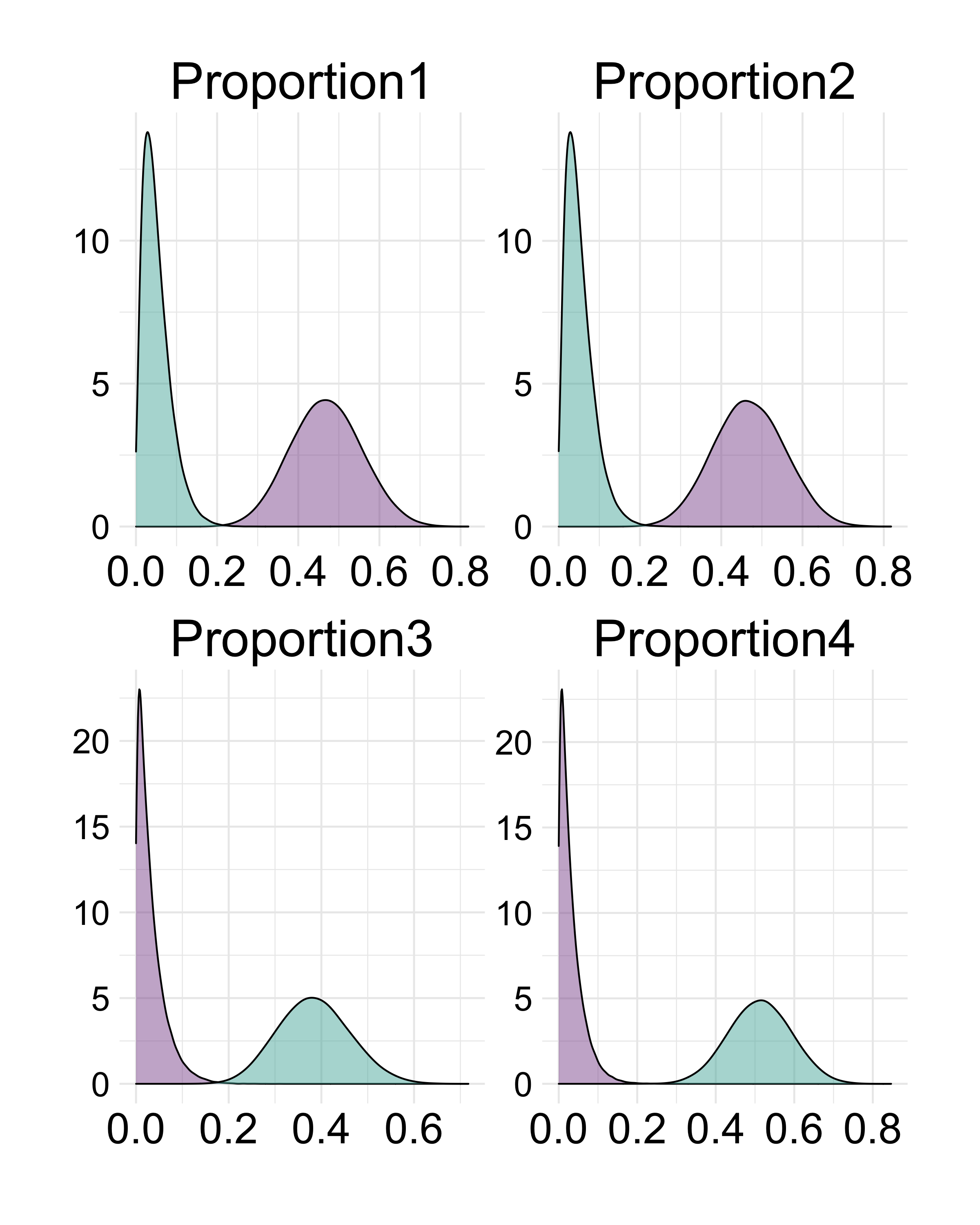}\hfill%
	\includegraphics[width=.32\textwidth,height=7cm]{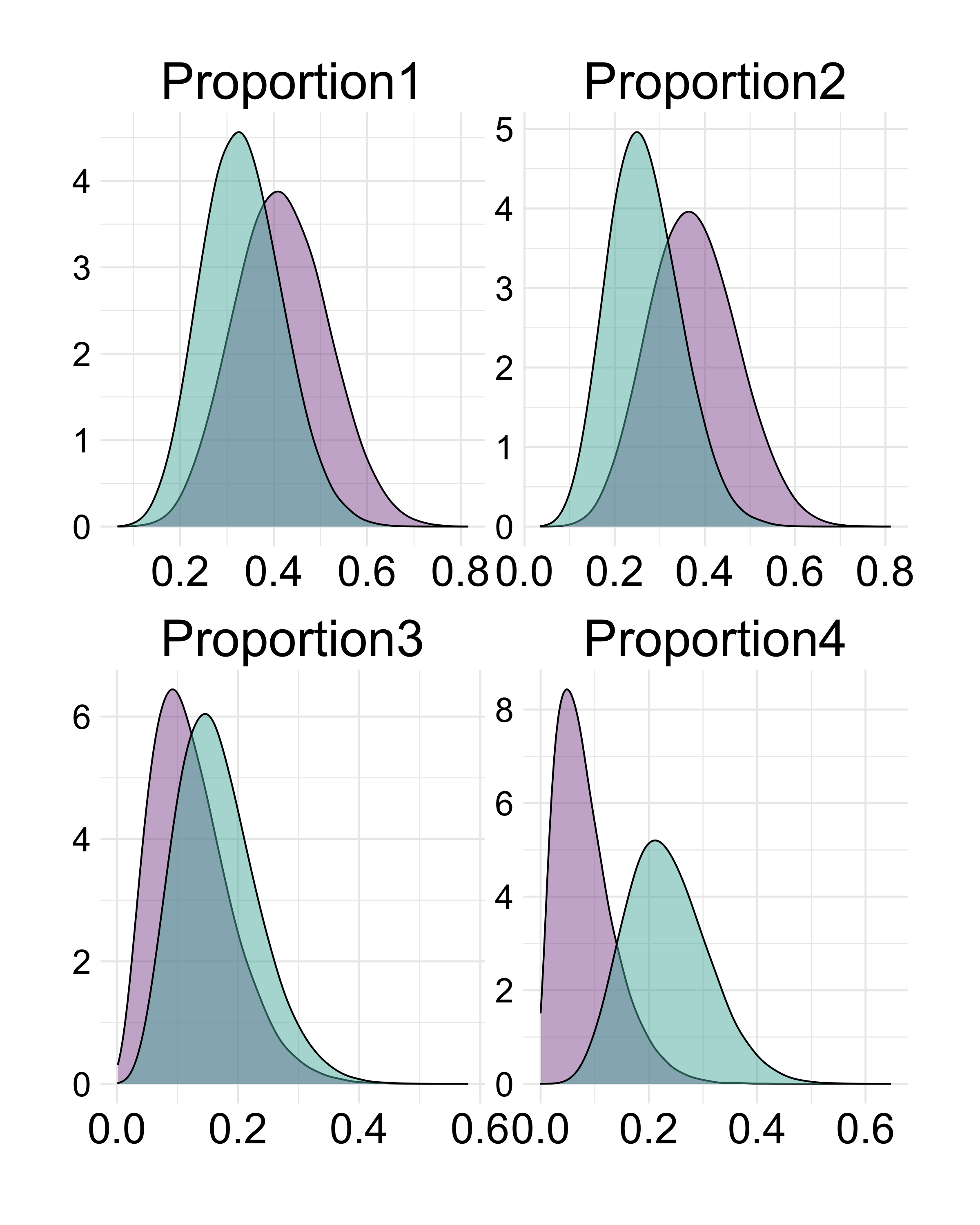}\hfill%
	\includegraphics[width=.32\textwidth,height=7cm]{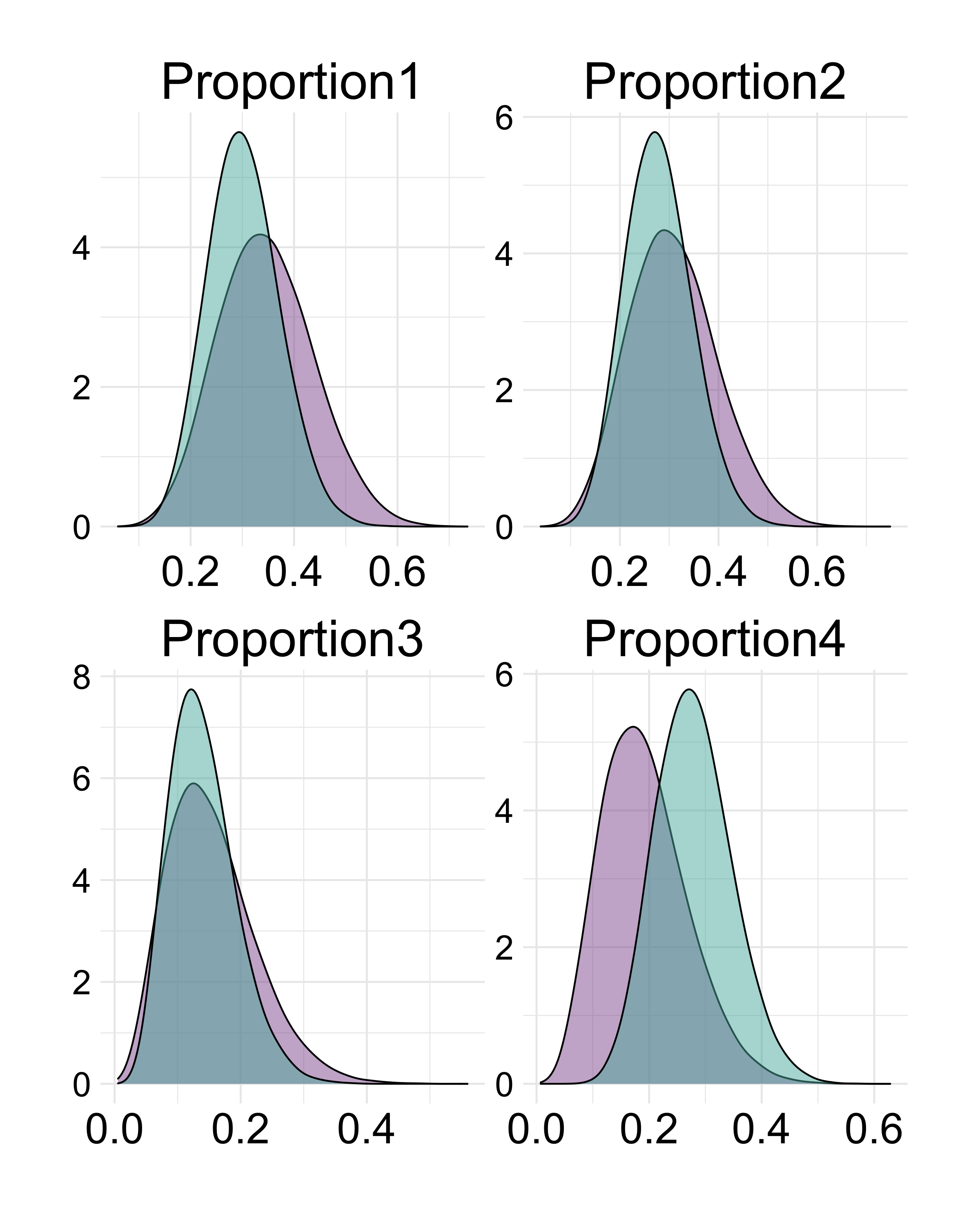}\hfill%
	\caption{Marginal densities of Dirichlet distributed data of two clusters with high (left), moderate (middle) and low (right) separation. Purple plots correspond to the first cluster ($\boldsymbol{\rho}_1$) and blue plots to the second  ($\boldsymbol{\rho}_2$). }\label{sim_plots}
\end{figure}

Configurations of high, moderate and low separation of the two component case are illustrated in Figure \ref{sim_plots}. The four panels show the marginal densities of proportions one to four, where colors correspond to the different Dirichlet components. In the left-side plots, the high separation case is displayed, where notably there is negligible overlap among clusters for all the proportions. With a Hellinger's distance of 0.75, the moderate scenario in the middle window shows some overlap in the four dimensions, with the last proportion being the most distinguishable between groups. A low separation setting is given on the right hand side of Figure \ref{sim_plots}, where the Hellinger's distance between the group-specific Dirichlet distributions is around 0.45. An increase in the overlapping area is observed in contrast with the moderate scenario, and the difference between the groups is now very subtle.

Table \ref{table_sim} shows that, in the two cluster setting, the partition used to simulate the data was completely recovered for both sample sizes as shown by the confusion matrix. Some confusion starts to arise as we increase the similarity between the two groups in the moderate scenario. Under $n=30$ the cluster-specific accuracy is 100\% for group 1 and 60\% for group 2. Evidently, the results depend on the sample size, where better performance is achieved for the same parameter configuration when $n$ is 50: the overall accuracy increased from 80\% to 94\% and the cluster-specific accuracy of group 2 went from 60\% to 88\%. Our third 2-cluster setting indicates that the amount of overlap characterized by a Hellinger's distance of approximately 0.45 is not well supported for the small sample sizes of 15 or 25 observations per group. It is important to bear these results in mind when working with our data application, where we should not expect to clearly identify subtle distinctions between clusters for the given sample sizes.

When three clusters are simulated, we denote as highly separated a scenario where the overlap among the three pairs is negligible, all with a Hellinger's distance approaching one. A moderate separation now refers to the case where one pair of cluster-specific distributions shows overlap. This is introduced for clusters 2 and 3, where the Hellinger's distance between this pair is around 0.72. In the low separation case, there is overlap between the three groups, all with pairwise Hellinger distances below 0.7. Results due to the 3-cluster setting displayed in the bottom-half of Table \ref{table_sim} show that, even with a reduced number of observations per group, the true partition is completely recovered if there is a high separation. One vanishing component is found under $n=30$ (moderate), in which case the most similar pair is merged. Although with some confusion, increasing the sample size enables the identification of three groups. This is similar for the low separation setting where there is approximately a single cluster under the smallest sample size but the original number of groups is spotted when $n$ is 50.

The analysis reported in Table \ref{table_sim} focused mainly on the inspection of the cluster partition where the MAP Dirichlet parameters associated to $\widetilde{\boldsymbol{z}}$ were displayed to contrast overall characteristics of the original and estimated groups. A careful investigation of parameter estimation under our model should reflect the uncertainty associated with those point estimates. For this purpose, 95\% posterior credible intervals are reported in Table \ref{table_post_high} for some settings. This output is readily provided by our Bayesian inference and shows that the true parameter values are all comprised by the intervals. In other words, the true values used to generate the data are likely under the posterior model, as expected.

\begin{table}[]
\centering
\small
\begin{tabular}{@{}lccccc@{}}
\toprule
\textbf{Scenario}                                                                    & \textbf{True parameters} & $\boldsymbol{\rho}_{k1}$   & $\boldsymbol{\rho}_{k2}$   & $\boldsymbol{\rho}_{k3}$   & $\boldsymbol{\rho}_{k4}$   \\ \midrule
\multirow{2}{*}{\begin{tabular}[c]{@{}l@{}}2 clusters\\ n = 50\end{tabular}} & (15, 15, 1, 1)     & 11.03 - 19.75 & 12.18 - 21.73 & 0.88 -1.60    & 0.95 - 1.74   \\
                                                                             & (2, 2, 15, 20)     & 1.45 - 2.64   & 1.70 - 3.07    & 11.55 - 20.39 & 13.77 - 24.31 \\ \cmidrule(l){3-6} 
\multirow{3}{*}{\begin{tabular}[c]{@{}l@{}}3 clusters\\ n = 50\end{tabular}} & (10, 12, 1, 0.5)     & 8.03 - 16.22  & 9.14 - 18.4   & 0.76 - 1.58   & 0.38 - 0.80   \\
                                                                             & (1, 2, 15, 18)     & 0.73 - 1.50   & 1.29 - 2.61   & 9.61 - 19.16  & 10.52 - 21.10 \\
                                                                             & (10, 10, 10, 10)     & 5.87 - 11.77  & 6.60 - 13.13  & 5.83 - 11.64  & 7.87 - 15.65  \\ \midrule
\end{tabular}
\caption{Posterior 95\% credible intervals for Dirichlet cluster-specific parameters. Results are obtained with the model fitted to the synthetic data sets of high separation with $n=50$.}\label{table_post_high}
\end{table}

In this section, we assessed the ability of our model to recover clusters of compositional data through simulation studies with configurations chosen to approach our motivating application. Data was drawn from 4-dimensional Dirichlet mixtures with two and three components varying distribution overlap. When increasing the number of clusters, the sample size was kept to investigate the sort of separation that is supported under a decreasing number of observations per group. Our model demonstrated being able to recover the true data partition even under small sample sizes when group separation is high. Introducing distribution overlap highlighted that the ability to capture moderate distinctions is dependent on the sample size. Further, 95\% credible intervals illustrated that the posterior distributions of model parameters are consistent with the data true parameter configurations.

\subsection{Model selection}\label{model_selection}

Under our approach, the number of groups is inferred using a model choice strategy. In this section, three alternative information criteria to guide the choice of $k$ are presented. Afterwards, our goal is to evaluate their performance for small sample sizes under controlled settings.

The ICL criterion was developed for mixture models and takes the missing data $\boldsymbol{z}$ into account. A non asymptotic version named exact ICL (\cite{ICL_exact}) can be obtained if the observed data likelihood is available, see also \cite{bertoletti2015}. Since it is not straightforward to integrate the cluster-specific parameters in our model, the approximate version of ICL is considered. The asymptotic ICL is based on the premise that if we can factorize $f(\underline{\boldsymbol{p}}, \boldsymbol{z}| \boldsymbol{\underline{\rho}}) = f(\underline{\boldsymbol{p}}|\boldsymbol{z},\boldsymbol{\underline{\rho}}) f(\boldsymbol{z}|\delta)$, then a BIC approximation is valid for $f(\underline{\boldsymbol{p}}| \boldsymbol{z},\boldsymbol{\underline{\rho}})$. The expression for $f(\boldsymbol{z}|\delta)$ is available in closed form and the approximation for $f(\underline{\boldsymbol{p}}|\boldsymbol{z}, \boldsymbol{\underline{\rho}})$ is

\begin{eqnarray*}
\log f(\underline{\boldsymbol{p}}|\boldsymbol{z}, \boldsymbol{\underline{\rho}}) \approx \max_{\boldsymbol{\underline{\rho}}} \log f(\underline{\boldsymbol{p}}|\boldsymbol{z},\boldsymbol{\underline{\rho}}) - \frac{\lambda_k}{2} \log n,
\end{eqnarray*}
where $k$ is the number of mixture components and $\lambda_{k}$ the number of free components in $\boldsymbol{\underline{\rho}}$. The proposed approximation is
where $k$ is the number of mixture components and $\lambda_{k}$ the number of free components in $\boldsymbol{\underline{\rho}}$. The proposed approximation is

\begin{eqnarray}\label{ICLas}
\mbox{ICL}(k) =  \log f(\underline{\boldsymbol{p}}|\boldsymbol{\widetilde{z}}, \boldsymbol{\underline{\boldsymbol{\widetilde{\rho}}}}) - \frac{\lambda_{k}}{2}\log n  + \log f(\boldsymbol{\widetilde{z}}|\delta),
\end{eqnarray}
where $(\boldsymbol{\widetilde{z}}, \boldsymbol{\underline{\boldsymbol{\widetilde{\rho}}}})$ denotes the joint MAP estimator of ($\boldsymbol{z}, \boldsymbol{\underline{\rho}}$) approximated from the MCMC sample.  

Various versions of the popular Deviance Information Criteria (DIC) were introduced for missing data models by \cite{dic}. Variants in this paper that are based on the observed likelihood cannot be easily computed for our model, so we chose to evaluate DIC$_5$, which depends on the conditional likelihood function. It is defined as

\begin{eqnarray}\label{dic5}
\mbox{DIC}_5 = -4E_{\boldsymbol{\underline{\rho}}, \boldsymbol{z}}[\log f(\underline{\boldsymbol{p}}, \boldsymbol{z}| \boldsymbol{\underline{\rho}})|\underline{\boldsymbol{p}}] + 2\log f(\underline{\boldsymbol{p}}, \boldsymbol{\widetilde{z}}| \boldsymbol{\underline{\boldsymbol{\widetilde{\rho}}}}), 
\end{eqnarray}
with expectations being estimated via Monte Carlo averages. 

ICL, DIC$_5$ and BIC were compared via simulation studies where we considered fitting the model with $k=1,\cdots,5$ for the eight data sets of high and moderate separations described in Section \ref{simulation}. In each scenario, the largest ICL and minimum values of BIC and DIC$_5$ indicate the preferred model choice. The three criteria were in agreement under highly separated clusters, selecting the $k$ value that was used to generate the data. Under moderate separation, a smaller $k$ was favoured twice by ICL and DIC$_5$, whilst it was overestimated by BIC. Detailed results can be found in the supplementary material. In conclusion, the three information criteria demonstrated similar performance for synthetic small sample data sets with the caveat that ICL and DIC$_5$ can underestimate $k$ when group separation is not as high, while overestimation can occur for BIC. We recommend that the agreement between ICL, DIC$_{5}$ and BIC is investigated in practical applications and, in case of divergence, the different model outputs should be compared.  


\section{Application to coral reefs}\label{application}
  
 In this section, we apply the proposed methodology to the data sets introduced in Section \ref{data_sets}. Models are fitted separately to each time point with various $k$, where the preferred number of clusters is indicated via the information criteria described in Section \ref{model_selection}. For each model, five parallel chains are run from different starting points for a total of 2.5M iterations. The first 500K are discarded as burn-in and $10,000$ thinned samples are used for inference. Label-switching is tackled in a post-processing step and the Brooks-Gelman-Rubin (BGR) statistic (\cite{bgr1}) is calculated for the relabelled Dirichlet parameters chains to assess convergence. The proposal variance parameters $\sigma_{\alpha}$ and $p_{var}$ were calibrated via pilot runs targeting acceptance rates for $\alpha$ and the elements of $\boldsymbol{\underline{\rho}}$ in the $30\%$ to $50\%$ range. We found the values $\sigma_{\alpha} = 0.5$ and $p_{var} = 0.7$ to produce good results in this application.  Additionally, we set $\delta = 1/2$, $\gamma = 0.2$, $\phi = 5$ and $\lambda =6$. 
 
\subsection{2012 ecological survey}\label{application_2012}

Results for the 2012 ecological survey of community composition are reported in this section. The model is fitted with $k=1, \cdots, 10$ and we inspect ICL, BIC and DIC$_5$. This is reported in the supplementary material, where the four-component Dirichlet mixture is suggested by the three criteria. As before, the BGR statistic was computed for the $\underline{\boldsymbol{\rho}}$ parameters and used to assess convergence. Having obtained values close to one and below 1.1, we combine the draws from five parallel chains and report inference from the resulting $50,000$ posterior samples due to the four cluster model fit.

The median and 95\% credible intervals (in parentheses) for the $\rho_{il}$ distributions are given in Table \ref{median_rho_2012}. To interpret these in terms of relative abundances, the statistics are also reported for the parameter vectors normalised by their sum (in bold). This indicates the presence of relative abundance patterns that can be characterized as follows. Group 1 (Algae/HC) is equally dominated by the presence of Algae and Hard coral (HC) estimated to be around 43\% of the total cover area, where the representation of Sand and Soft coral is low. The main feature of the second group (Algae/HC/SC) is a higher than average abundance of Soft coral, in equal abundance with Algae and Hard coral. In contrast, the third cluster (HC/Sand) is characterized by a higher proportion of Sand along with the presence of Hard corals. Finally, group 4 (HC mix) is characterized by the dominance of Hard coral with around 20\% median abundance of Algae, Sand and HC. Given that the proportion of HC tends to be relatively constant among all clusters (between 0.30-0.40), we assume that the four species are reasonably well represented in this last group. HC mix is the most represented cluster, with a total number of 20 transects followed by Algae/HC (16 transects), Algae/HC/SC (12 transects) and finally HC/Sand (10 transects).

Additionally, the posterior entropy distributions of cluster-specific parameter vectors are investigated. Draws from the cluster-specific parameters obtained from the Bayesian model output are used to calculate the posterior entropy distribution for each group with their 5\%, 50\% and 95\% quantiles reported in Table \ref{entropy_2012}. This analysis allows us to explore the within group variability and tells us that there is less uncertainty associated with the distributions of groups two and one. Accordingly, there is a higher species diversity in the HC mix cluster.

\begin{figure}[H]
	\centering
		\includegraphics[width=.5\textwidth]{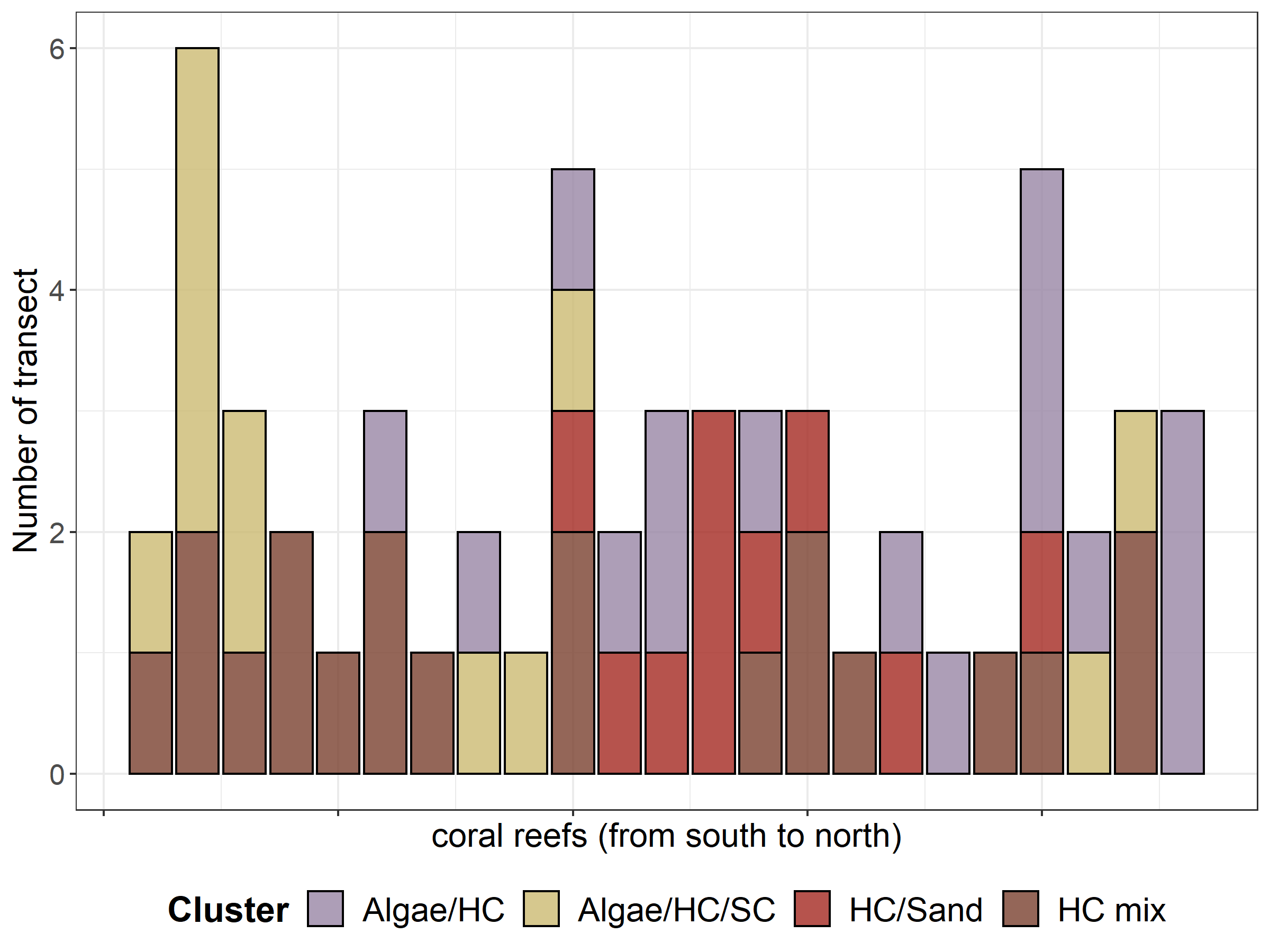}
		\captionof{figure}{Spatial distributions of clusters within coral reefs sorted from south to north, for the year 2012. HC and SC denote "Hard coral" and "Soft coral", respectively.}\label{clust.spat}
\end{figure}

Spatial distributions of the 2012 clusters are illustrated in Figure \ref{clust.spat} and reveal a strong intra-reef heterogeneity with only three reefs for which more than one transect was grouped within the same cluster. The cluster HC/Sand was estimated for a few consecutive reefs between Ribbon-5 reef and Tijou reef, a part of the Great Barrier Reef well characterized by its strong currents (\cite{colberg2020high}). Overall, the HC mix cluster tends to be located on the side of the reef exposed to oceanic waves and Algae/HC and Algae/HC/SC on the north or south ends of the reef, which are more influenced by the tidal currents.     

\renewcommand{\arraystretch}{1}
\begin{table}
		\centering
		\small
\begin{tabular}{ccccc}
\hline
\textbf{} & \textbf{Algae}                                                                                       & \textbf{Hard coral}                                                                                  & \textbf{Sand}                                                                                        & \textbf{Soft coral}                                                            \\ \hline
Group 1   & {\color[HTML]{000000} \begin{tabular}[c]{@{}c@{}}9.63 (5.01 - 19.32)\\ \textbf{0.43 (0.36-0.5)}\end{tabular}} & \begin{tabular}[c]{@{}c@{}}9.83 (5.45 - 17.54)\\ \textbf{0.43 (0.38-0.49)}\end{tabular}                       &  \begin{tabular}[c]{@{}c@{}}0.64 (0.39 - 0.98)\\ \textbf{0.03 (0.01-0.05)}\end{tabular} & \begin{tabular}[c]{@{}c@{}}2.34 (1.44 - 3.94)\\ \textbf{0.10 (0.07-0.15)}\end{tabular}  \vspace{0.2cm} \\
Group 2   & \begin{tabular}[c]{@{}c@{}}8.54 (3.43 - 16.75)\\ \textbf{0.28 (0.2-0.35)}\end{tabular}                        &  \begin{tabular}[c]{@{}c@{}}9.56 (4.07 - 16.89)\\ \textbf{0.31 (0.23-0.4)}\end{tabular} & \begin{tabular}[c]{@{}c@{}}2.08 (1.01 - 4.57)\\ \textbf{0.07 (0.03-0.22)}\end{tabular}                        & \begin{tabular}[c]{@{}c@{}}10.55 (3.2 - 20.13)\\ \textbf{0.34 (0.19-0.41)}\end{tabular} \vspace{0.2cm} \\
Group 3   & \begin{tabular}[c]{@{}c@{}}4.64 (2.38 - 8.12)\\ \textbf{0.18 (0.12-0.26)}\end{tabular}                        & \begin{tabular}[c]{@{}c@{}}9.65 (4.33 - 16.76)\\ \textbf{0.37 (0.25-0.44)}\end{tabular}                       & \begin{tabular}[c]{@{}c@{}}9.97 (3.45 - 19.38)\\ \textbf{0.38 (0.2-0.48)}\end{tabular}                        & \begin{tabular}[c]{@{}c@{}}1.99 (1.05 - 4.38)\\ \textbf{0.07 (0.04-0.22)}\end{tabular} \vspace{0.2cm} \\
Group 4   & \begin{tabular}[c]{@{}c@{}}5.35 (3.16 - 8.64)\\ \textbf{0.21 (0.17-0.26)}\end{tabular}                        & \begin{tabular}[c]{@{}c@{}}11.31 (6.36 - 19.03)\\ \textbf{0.44 (0.39-0.49)}\end{tabular}                      & \begin{tabular}[c]{@{}c@{}}4.41 (2.29 - 7.77)\\ \textbf{0.17 (0.12-0.25)}\end{tabular}                        & \begin{tabular}[c]{@{}c@{}}4.41 (2.06 - 8.41)\\ \textbf{0.17 (0.12-0.24)}\end{tabular}  \\ \hline
\end{tabular}
		\captionof{table}{Median and 95\% credible interval of posterior Dirichlet parameters of the 2012 data set. In bold, the cluster parameters normalised by their sum.}\label{median_rho_2012}
\end{table}

\begin{table}[]
    \centering
    \small
\begin{tabular}{@{}cccc@{}}
		\toprule
		\textbf{} & \textbf{Q5} & \textbf{Median} & \textbf{Q95}  \\ \midrule
		Group 1   & $-329.64$ & $-181.97$ & $-102.14$  \\
		Group 2   & $-333.91$ & $-194.65$ & $-100.39$  \\
		Group 3   & $-276.28$ & $-167.52$ &  $-93.52$     \\
		Group 4   & $-233.45$ & $-148.47$ &  $-92.49$      \\ \bottomrule
	\end{tabular}
	\captionof{table}{5\%, 50\% and 95\% quantiles of the groups posterior entropy for the 2012 data set.}\label{entropy_2012}
\end{table}
	
\subsection{Clustering over the years}

Our Bayesian latent allocation model was applied to the compositional data collected in 2014, 2016 and 2017. We fitted the model with $k=1, \cdots, 10$, for which the information criteria are provided in the supplement. In 2014, ICL and DIC$_5$ indicate $k=1$ while $k=2$ is suggested by BIC. For the 2016 and 2017 data, $k=2$ is chosen according to the three criteria. We present the results due to $k=2$ in 2014 to investigate continuity of these two clusters. In Table \ref{clusters_years}, it can be seen that, for each year, one cluster is composed of a more balanced abundance of the four communities and the other is dominated by HC and Algae. These clusters are identical to HC mix and Algae/HC which were the two most represented in the 2012 data.

\begin{table}[H]
\centering
\small
\begin{tabular}{@{}lcccccc@{}}
\toprule
           & \multicolumn{2}{c}{\textbf{2014}}                                                                                          & \multicolumn{2}{c}{\textbf{2016}}                                                                                           & \multicolumn{2}{c}{\textbf{2017}}                                                                                          \\ \midrule
           & \begin{tabular}[c]{@{}c@{}}Group 1\\ ($n = 7$)\end{tabular} & \begin{tabular}[c]{@{}c@{}}Group 2\\ ($n = 26$)\end{tabular} & \begin{tabular}[c]{@{}c@{}}Group 1\\ ($n = 12$)\end{tabular} & \begin{tabular}[c]{@{}c@{}}Group 2\\ ($n = 37$)\end{tabular} & \begin{tabular}[c]{@{}c@{}}Group 1\\ ($n = 7$)\end{tabular} & \begin{tabular}[c]{@{}c@{}}Group 2\\ ($n = 28$)\end{tabular} \\ \midrule
Algae      & \begin{tabular}[c]{@{}c@{}}0.21 \\ (0.16-0.25)\end{tabular} & \begin{tabular}[c]{@{}c@{}}0.37 \\ (0.30-0.45)\end{tabular}  & \begin{tabular}[c]{@{}c@{}}0.2 \\ (0.16-0.25)\end{tabular}   & \begin{tabular}[c]{@{}c@{}}0.41 \\ (0.36-0.45)\end{tabular}  & \begin{tabular}[c]{@{}c@{}}0.23 \\ (0.15-0.31)\end{tabular} & \begin{tabular}[c]{@{}c@{}}0.43 \\ (0.38-0.47)\end{tabular}  \\
Hard coral & \begin{tabular}[c]{@{}c@{}}0.42 \\ (0.37-0.46)\end{tabular} & \begin{tabular}[c]{@{}c@{}}0.41 \\ (0.36-0.46)\end{tabular}  & \begin{tabular}[c]{@{}c@{}}0.39\\ (0.32-0.45)\end{tabular}   & \begin{tabular}[c]{@{}c@{}}0.41 \\ (0.38-0.45)\end{tabular}  & \begin{tabular}[c]{@{}c@{}}0.38 \\ (0.27-0.47)\end{tabular} & \begin{tabular}[c]{@{}c@{}}0.38 \\ (0.33-0.42)\end{tabular}  \\
Sand       & \begin{tabular}[c]{@{}c@{}}0.24 \\ (0.17-0.32)\end{tabular} & \begin{tabular}[c]{@{}c@{}}0.05 \\ (0.03-0.08)\end{tabular}  & \begin{tabular}[c]{@{}c@{}}0.28 \\ 0.21-0.35)\end{tabular}   & \begin{tabular}[c]{@{}c@{}}0.03 \\ (0.02-0.05)\end{tabular}  & \begin{tabular}[c]{@{}c@{}}0.28\\ (0.18-0.38)\end{tabular}  & \begin{tabular}[c]{@{}c@{}}0.04 \\ (0.03-0.06)\end{tabular}  \\
Soft coral & \begin{tabular}[c]{@{}c@{}}0.14 \\ (0.09-0.18)\end{tabular} & \begin{tabular}[c]{@{}c@{}}0.17 \\ (0.12-0.21)\end{tabular}  & \begin{tabular}[c]{@{}c@{}}0.13 \\ (0.09-0.17)\end{tabular}  & \begin{tabular}[c]{@{}c@{}}0.15 \\ (0.12-0.17)\end{tabular}  & \begin{tabular}[c]{@{}c@{}}0.12 \\ (0.07-0.18)\end{tabular} & \begin{tabular}[c]{@{}c@{}}0.16 \\ (0.13-0.19)\end{tabular}  \\ \bottomrule
\end{tabular}
\caption{Normalised posterior median and 95\% credible interval (in parentheses) of Dirichlet parameters for the models selected in 2014, 2016 and 2017.}\label{clusters_years}
\end{table}

\begin{figure}[H]
	\centering
	\begin{minipage}[b]{.35\textwidth}
		
		\includegraphics[width=\textwidth]{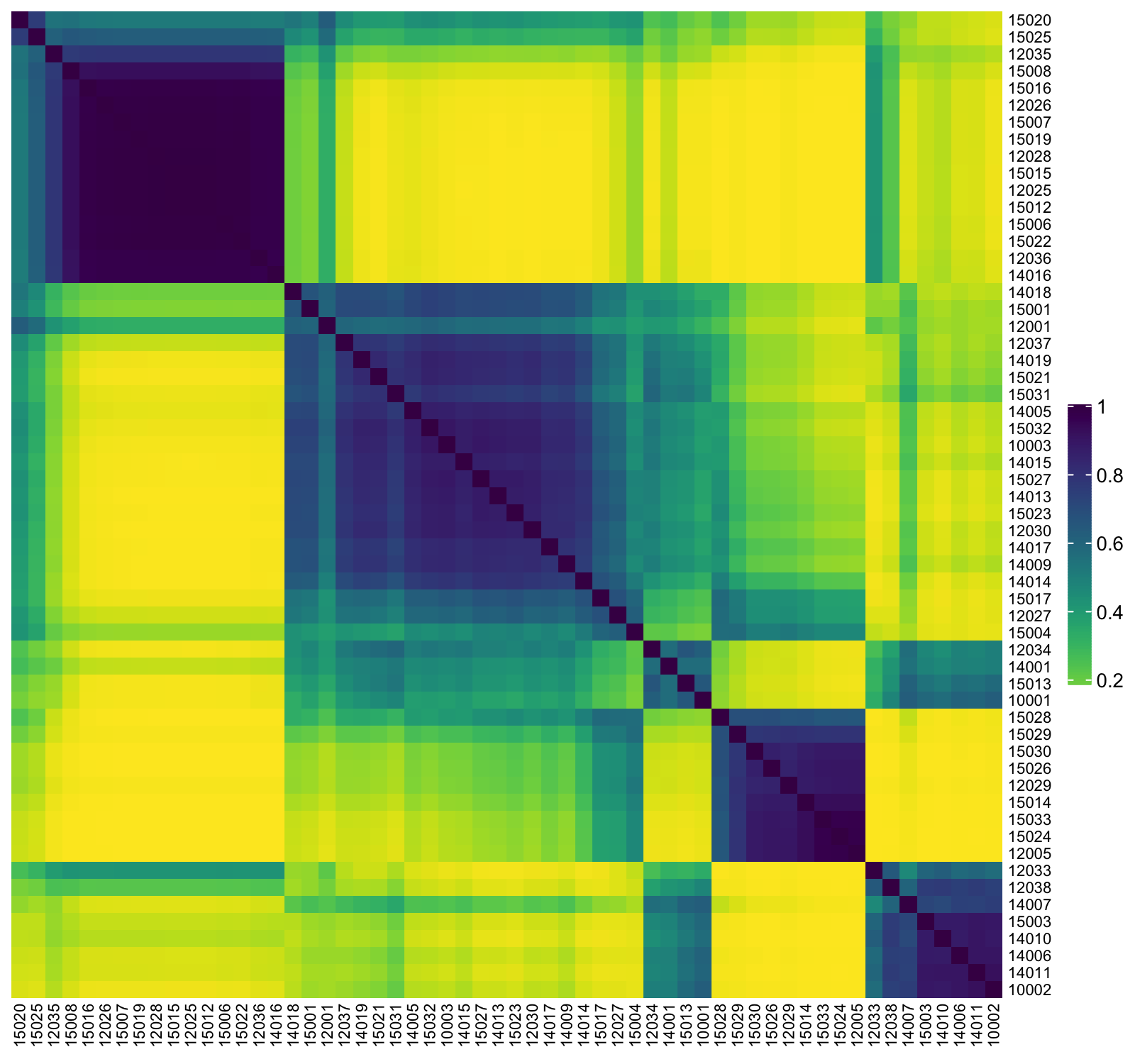}
		\includegraphics[width=\textwidth]{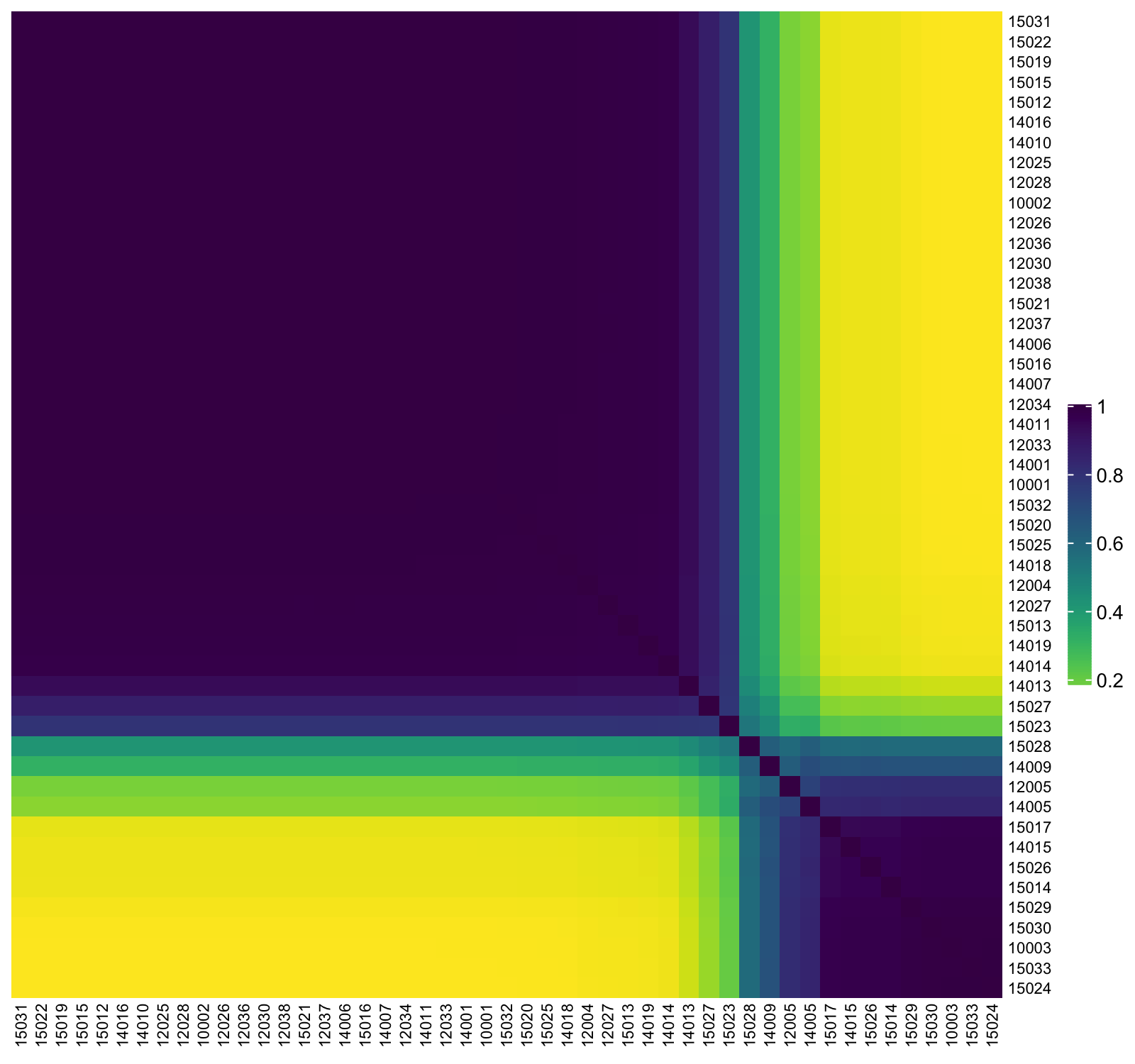}
		
	\end{minipage}
	\begin{minipage}[b]{.35\textwidth}
		
	\includegraphics[width=\textwidth]{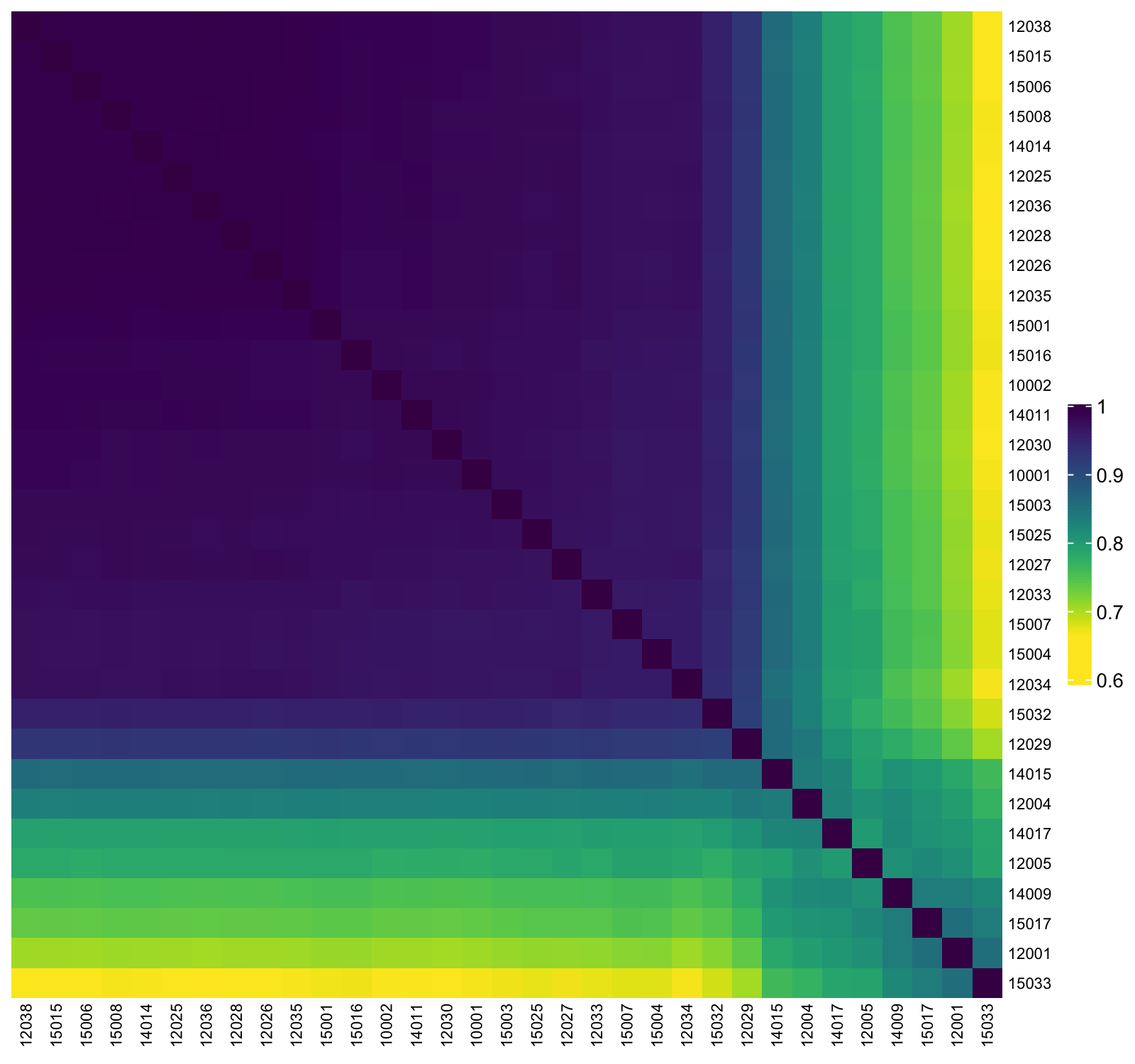}
		\includegraphics[width=\textwidth]{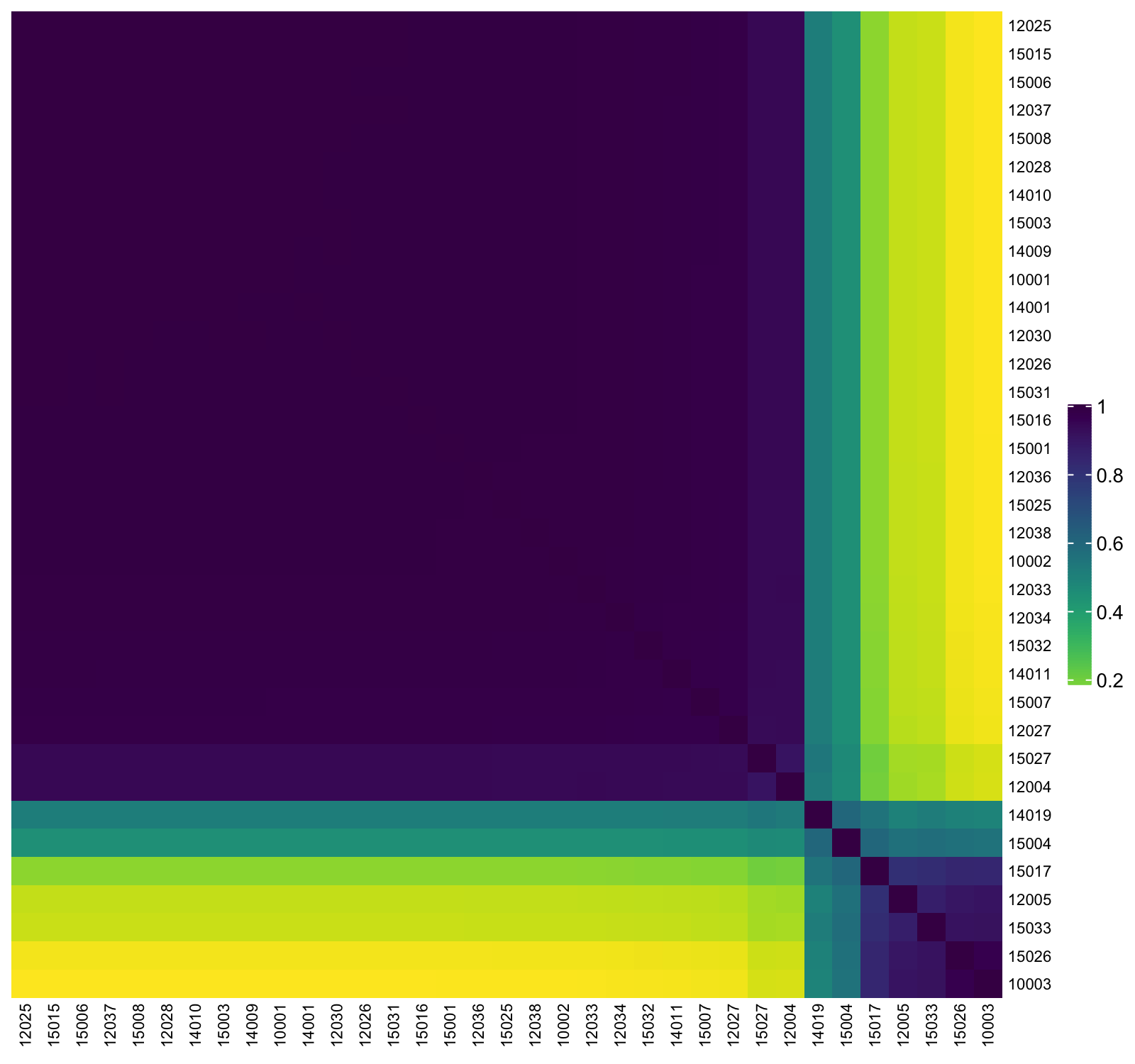}

	\end{minipage}
\caption{Heatmaps of posterior probability matrices of 2012 (top-left), 2014 (top-right), 2016 (bottom-left) and 2017 (bottom-right). A high probability (in blue colour) means that pair of transects has high chance to be allocated to the same cluster.}\label{heatmaps}
\end{figure}

Despite these similarities, pairwise comparisons of posterior probabilities that two transects are allocated within the same cluster are gradually increasing through the years (Figure~ \ref{heatmaps}). We find that only 5.4\% of the pairwise probabilities were above 0.9 in 2012. The similarity increases among the years, where  56.8\%, 50.76\% and 64.20\% of pairs have at least 0.9 probability of being allocated to the same group in 2014, 2016 and 2017, respectively.

\begin{figure}[H]
\centering
 \includegraphics[width=0.7\textwidth]{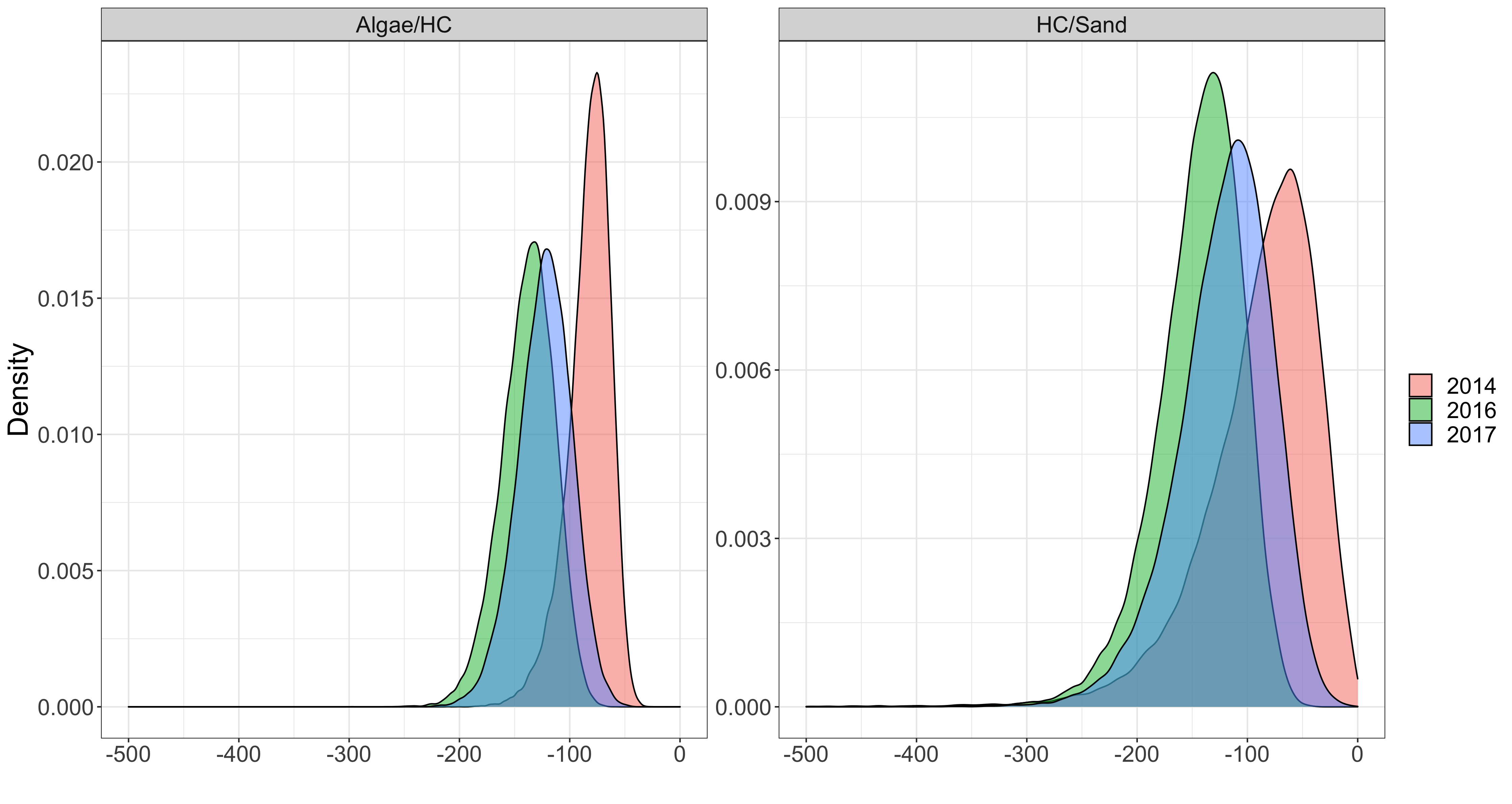}
 \caption{Posterior entropy distribution of the Algae/HC and HC/Sand clusters in 2014, 2016 and 2017.} \label{entropy_years}
\end{figure}

The decrease in the number of clusters found from 2014 onwards naturally indicates a reduction in diversity of composition patterns as do the increased pairwise cluster membership probabilities in Figure \ref{heatmaps}. Although the same number of groups is found in 2014, 2016 and 2017, we can assess the evolution of reef compositions throughout these years by comparing their within-cluster diversity. To this end, the posterior entropy distributions of the HC mix and Algae/HC clusters over this period are calculated and displayed in Figure \ref{entropy_years}. Within both clusters, we observe a decrease in entropy of the two most recent years in comparison to 2014. Although it seems to increase slightly in the final year for both groups, the distribution is still concentrated at lower values than compared to 2012. This indicates a reduced diversity of compositions inside the HC mix and Algae/HC clusters in 2016, which persists in 2017.

\section{Conclusions and future research}

Ecological surveys conducted over four years in the GBR gave rise to species composition data at various reef localities. The aim of this paper was to provide a better understanding of composition patterns and how these were affected by a unprecedented sequence of disturbance events. To achieve this, a model for clustering compositional data in the simplex was introduced.

The proposed model dictates that observations arise from a finite mixture of Dirichlet distributions with latent cluster memberships. Conditionally on the unobserved allocations, the model likelihood assumes a simple form thanks to the chosen hierarchical specification. Markov Chain Monte Carlo methods were outlined to perform Bayesian inference by sampling from the posterior distribution of model parameters and latent variables, demonstrating an excellent performance in controlled scenarios. Computational efficiency, label switching, and model choice aspects were also addressed. The model fit to the 2012 GBR ecological survey enabled the identification of wave exposure associated to the reef geographical location as a factor affecting the relative species abundances. In addition to the immediate ecological inferences arising from these results, this finding has crucial implications for the design of future ecological surveys conducted on the GBR. For example, it is necessary to include transects from north to south in order to have an accurate representation of the entire ecosystem. Fitting the proposed model to the ecological surveys conducted over four years showed growing similarity of compositions of the surveyed reef locations, as evidenced by the decrease in the number of clusters and increased pairwise similarity of observations.

Some extensions of the proposed compositional data clustering model that we hope to address are the following. Covariates can be included via a regression structure for the Dirichlet parameters with, for example, a logarithm link function. By allowing the regression parameters to be group and composition specific, a practitioner can evaluate if covariates of interest have an effect on compositions for distinct groups. Another possible model extension is considering a generalised Dirichlet likelihood. This distribution provides a more flexible covariance structure, at the cost of having almost twice the number of parameters than the standard Dirichlet. Finally, the results from our application motivate a clustering model that takes into account the spatial location of the reef communities as well as the temporal evolution of the coral reef compositions.

\section*{Acknowledgements} This publication has emanated from research conducted with the financial support of Science Foundation Ireland under Grant number 18/CRT/6049.  
The Insight Centre for Data Analytics is supported by Science Foundation Ireland under Grant Number 12/RC/2289$\_$P2. Support from the Australian Research Council Centre of Excellence in Mathematical Frontiers (KM, JV) and the ARC Laureate Fellowship Program (KM) is also acknowledged.

\section*{Code and data} The data sets and Python/R code to fit the proposed model are available online at \url{https://github.com/luizapiancastelli/Reef_Clustering}.

\bibliographystyle{apalike} 
{\footnotesize\bibliography{refs.bib}       }

\end{document}